\documentclass[trackchanges,twocolumn]{aastex7}

 \usepackage{url}
\usepackage{graphicx}   
\usepackage{amsmath}    
\usepackage{amssymb}    

\makeatletter
\newcommand*{\hyperlinkcite}[1]{\hyper@link{cite}{cite.#1}}

\newcommand{\OmK}{\Omega_\mathrm{K}}

\newcommand{\gameff}{\gamma_{\mathrm{eff}}}
\makeatother

\newcommand{\PK}{\hyperlinkcite{paardekooper2011}{P11}}

\begin{document}

\title{Planet-Disk Interactions and the Convective Overstability. I. Low Mass Planets}

\author[0000-0002-0496-3539]{Marius Lehmann}
\email{mariusl@iastate.edu}
\affiliation{Department of Physics and Astronomy, Iowa State University, Ames, IA 50011, USA}
\affiliation{Institute of Astronomy and Astrophysics, Academia Sinica, Taipei 10617, Taiwan}

\author[0000-0002-8597-4386]{Min-Kai Lin}
\email{mklin@asiaa.sinica.edu.tw}
\affiliation{Institute of Astronomy and Astrophysics, Academia Sinica, Taipei 10617, Taiwan}
\affiliation{Physics Division, National Center for Theoretical Sciences, Taipei 10617, Taiwan}


\begin{abstract}
Rapid inward migration driven by Type I torques threatens the survival of low-mass planets in their nascent protoplanetary disks (PPDs). Positive co-rotation torques offer a potential solution, but require viscous diffusion to remain unsaturated. However, it is unclear if (magneto)-hydrodynamic turbulence provides the necessary diffusion, and disk profiles supporting such torques are often also susceptible to the Convective Overstability (COS) for suitable gas cooling timescales. 
To this end, we investigate torques on low-mass planets through radially global 2D (razor-thin) and vertically unstratified 3D hydrodynamic simulations of PPDs with thermal diffusion and optically thin cooling. Our 3D models with thermal diffusion, which allows COS development, show systematically different torque behavior compared to 2D models, wherein COS is absent. In 3D, the COS saturates into large-scale, long-lived vortices that migrate radially and interact gravitationally with the embedded planet. When these vortices encounter the planet, they typically provide positive torque "kicks" counteracting inward migration, as the less-massive vortices are scattered onto horseshoe orbits by the more-massive planet. We validate our simulation methods against the theoretical framework of Paardekooper et al. (2011) and demonstrate that COS-induced torque modifications can extend migration timescales by factors of approximately 10. For plausible disk models, our results suggest that COS activity can lengthen migration timescales sufficiently to overlap with, or even exceed Super-Earth formation windows (0.1-5 Myr). In contrast, simulations with optically thin cooling do not show significant torque modifications, as COS saturates in near-axisymmetric structures without producing large-scale vortices for the disk models considered here.
\end{abstract}




\section{Introduction}\label{sec:intro}

Planet migration in protoplanetary disks is a foundational process that dictates the final architecture of planetary systems \citep{paardekooper2023}. In the canonical model for low-mass planets, the gravitational interaction between the planet and the surrounding gaseous disk gives rise to Type I torques, which are expected to drive rapid, inward migration \citep{ward1997, tanaka2002}. The timescale for this migration can be significantly shorter than the disk's lifetime, presenting a formidable challenge to the survival of nascent planets and the formation of the observed population of exoplanets at moderate orbital separations.

For decades, turbulence driven by the magnetorotational instability (MRI) \citep{balbus1991} was considered the primary agent of angular momentum transport in disks and a key factor in modifying planet migration \citep{paardekooper2023}. However, a modern understanding of disk chemistry suggests that large portions of protoplanetary disks, particularly the dense, cool midplane regions where planets form, have ionization levels too low to sustain the MRI \citep{gammie1996, armitage2011, turner2014, cleeves2015}. The existence of these extensive ``dead zones'' has prompted a paradigm shift, elevating purely hydrodynamic instabilities from secondary phenomena to significant drivers of dynamical evolution in the most critical regions for planet formation \citep{lesur2023}. Since hyrodynamic instabilities have been found to form complex large scale flow structures, such as disk corrugation modes, zonal flows and vortices, this shift reframes the central problem. It is no longer sufficient to ask what value of effective viscosity (the $\alpha$ parameter) a certain type of turbulence amounts to. Instead, one must investigate how the specific nature of planet-disk interaction—that is, how the Lindblad and corotation torques that drive migration—is modified by the distinct flow structures governed by non-magnetic hydrodynamic processes \citep{paardekooper2023}. Among these, the Vertical Shear Instability (VSI: \citealt{urpin1998,urpin2003,nelson2013,barker2015,lin2015}) and the Convective Overstability (COS: \citealt{klahr2014,lyra2014,latter2016}]) have emerged as leading candidates.

Recent work has revealed a fundamental dichotomy in the nonlinear outcomes of the main hydrodynamic instabilities, with important consequences for planet migration.
The Vertical Shear Instability (VSI), driven by vertical shear in thermally stratified disks with short cooling times, typically saturates into vertically global turbulence \citep{nelson2013}.  While early global 3D simulations reported the spontaneous emergence of vortices and zonal flows \citep{manger2018, manger2020, manger2021, pfeil2020, lehmann2022}, more recent high-resolution models suggest that such coherent structures are transient or absent under simplified thermodynamics, yielding predominantly stochastic “fluffy’’ turbulence with $\alpha \sim 10^{-3}$ and no persistent large-scale features \citep{lesur2025}.  In such VSI-active disks, embedded low-mass planets experience highly variable but on average inward torques \citep{stoll2017}.

In contrast, recent simulations of the COS \citep{raettig2021,lehmann2024,lyra2024,lehmann2025} with similar resolution as used in \citet{lesur2025} do find the emergence of persistent radially large-scale flow structures. The COS is a form of oscillatory convection driven by a negative radial entropy gradient (for the usual case of a declining pressure profile), a condition that can be met in disk regions with steep temperature gradients or flat density profiles, 
paired with moderate gas cooling. The nonlinear saturation of the COS, observed in both 3D local and radially global simulations, is characterized by the formation of large-scale, coherent axisymmetric zonal flows and long-lived vortices \citep{lyra2014, lehmann2024}. 

It is, however, not clear how vortices affect the migration of nearby planets, especially those generated self-consistently by hydrodynamic instabilities. \citet{faure2016} explored this interaction in the special environment of a disk's dead zone inner edge. This interface is predicted to act as a powerful planet trap on its own \citep{masset2006}, where a sharp positive surface density gradient is expected to halt inward migration via a strong positive corotation torque. However, this same density maximum is unstable to the Rossby Wave Instability, leading to a cycle where large vortices are repeatedly generated, which then migrate inward. In their simulations, these vortices were an order of magnitude more massive than the embedded planet, allowing them to gravitationally capture the planet and drag it inward, thereby disrupting the otherwise stable trap. This leaves open the question of how migration is affected by less-massive vortices in regions without such a strong, pre-existing trapping structure—a scenario more typical for the dynamics in fully-developed hydrodynamic turbulence.

This study aims to investigate how disk-planet torques are influenced by turbulence self-consistently driven by the COS in three-dimensional, vertically unstratified protoplanetary disks. Previous related work has explored different facets of this complex environment. For instance, \citet{gomes2015} studied the evolution of planet-induced vortices, which are triggered by the Rossby Wave Instability (RWI; \citealt{lovelace1999,li2000,li01}) at the edges of a gap carved by a high-mass planet. Their two-dimensional simulations incorporated thermal relaxation and a negative radial buoyancy profile, conditions that permit the Sub-critical Baroclinic Instability to amplify vortices \citep{lesur2010}. However, their study differs from ours in two fundamental aspects: they focused on vortices generated by a massive, gap-opening planet, and their razor-thin 2D setup cannot capture the fully three-dimensional dynamics of the COS, which is suppressed in 2D. Our work addresses a distinct physical problem by investigating the torques on a low-mass, non-gap-opening planet embedded within fully developed 3D COS turbulence.

We perform global 2D and 3D hydrodynamic simulations of a PPD containing a low-mass, non-migrating planet embedded in a COS-active region. By fixing the planet's orbit, we can cleanly probe the torque landscape generated by the self-consistent interaction between the planet's gravitational potential and the COS-driven turbulence. This work serves as a necessary first step before tackling the more complex problems of migrating planets or the feedback from higher-mass planets. The differences between 2D and 3D simulations will directly reveal the effect of COS turbulence. 

The paper is organized as follows. Section \ref{sec:motivation} outlines the motivation for examining the influence of the COS on planet–disk interactions. In Section \ref{sec:pk11}, we revisit the theoretical framework describing torques acting on low-mass planets embedded in viscous, non-isothermal PPDs. Section \ref{sec:hydro} presents the hydrodynamic model and simulation setup. In Section \ref{sec:results2D}, we validate our simulations against analytical expectations. Section \ref{sec:results} contains the main results from our 2D and 3D simulations, highlighting the impact of COS on disk–planet torques. Finally, in Section \ref{sec:discussion}, we place our findings in a broader PPD context, discuss implications for planet migration, and outline key caveats.

\section{Theoretical motivation}\label{sec:motivation}

We further motivate the study of disk-planet interaction under the COS by noting the common thread between them: the disk's radial entropy profile, $\partial_rS$. Here, $r$ is the cylindrical radius, and 
\begin{align}\label{eq:ent}
S\equiv \ln{\left(\frac{P}{\rho^\gamma}\right)}
\end{align}
 is the dimensionless specific entropy, $\gamma$ is the gas adiabatic index, and $\rho$ and $P$ are the gas density and pressure, respectively. 

 In the idealized limit of an adiabatic, razor-thin power-law disk, the total non-linear torque $\Gamma$ acting on a low-mass planet is proportional to 
 \begin{align}\label{adiabatic_torque_formula}
     \Gamma \propto \underbrace{-\left(2.5 + 1.7q - 0.1s\right)}_{\text{Lindblad}} + \underbrace{1.1\left(\frac{3}{2} - s\right)}_\text{vortensity} + \underbrace{7.9\frac{\xi}{\gamma}}_\text{entropy},
 \end{align}
 where $-q$, $-s$, and $-\xi$ are power-law indices for the disk's radial temperature, surface density, and entropy profiles, respectively \citep{paardekooper2010}\footnote{Eq. \ref{adiabatic_torque_formula} specifically applies to a softening length of the planet potential equal to $0.4$ times the pressure scale-height, but this does not affect the signs of the torque contributions.} The first, second, and third terms represent contributions stemming from Lindblad torques, co-rotation torques related to the disk's vortensity profile (ratio of surface density to vorticity), and co-rotation torques related to the disk's entropy profile. Eq. \ref{adiabatic_torque_formula} indicates a negative entropy gradient ($\xi>0$) yields a corresponding \emph{positive} entropy-related co-rotation torque. This offers the possibility of slowing or reversing inward migration. 

However, disk regions with $\partial_r S < 0$, coupled with a typically negative pressure gradient, $\partial_r P < 0$, would yield an adverse radial buoyancy frequency profile, $N_r$, such that 
\begin{equation}\label{eq:nr2}
N_r^2  \equiv -\frac{1}{\gamma\rho}\frac{\partial P}{\partial r}\frac{\partial S}{\partial r} < 0 
\end{equation}
(see Section \ref{sec:hydro} for an explicit expression for power-law disks). When this condition is met \emph{and} the disk undergoes thermal losses, a fluid parcel undergoing epicylic motions can exchange heat with its surroundings such that its oscillation amplitude grows owing to buoyant accelerations --- this is the COS \citep{latter2016}. 

Taken together, this suggests that disk regions favorable for retaining protoplanets may also be convectively unstable. It is then natural to question how COS activity influences the entropy-related co-rotation torque. For example, in laminar disks, co-rotation torques eventually saturate or vanish as the co-orbital material undergoes phase mixing. However, if such regions are COS-active, could COS-driven turbulence de-saturate and sustain a positive co-rotation torque? 

Furthermore, if sufficiently vigorous, the COS saturates in the formation of zonal flows \citep{teed2021}, which subsequently break up into vortices \citep{lyra2014,lyra2024,lehmann2024,lehmann2025,teed2025}. These vortices are the main drivers of radial angular momentum transport and mass accretion driven by the COS. How do these vortices affect the torque on low mass planets, and consequently, their migration?

To address these issues, we develop simulations that permit both entropy-related torques and the COS. As the COS amounts to growing inertial waves, it is necessary to include the disk's vertical dimension (though not stratification) and thermal losses, the latter of which we model as optically thin cooling or radiative diffusion. For comparison with previous disk-planet studies, we also need to adapt semi-analytic torque formulas that account for dissipation \citep{paardekooper2011} (\PK~ hereafter), rather than using the adiabatic result represented by Eq. (\ref{adiabatic_torque_formula}).

\section{Planet-disk interactions in non-isothermal, viscous thin disks}\label{sec:pk11}

  \begin{figure*}
 \centering 
 	\includegraphics[width=\textwidth]{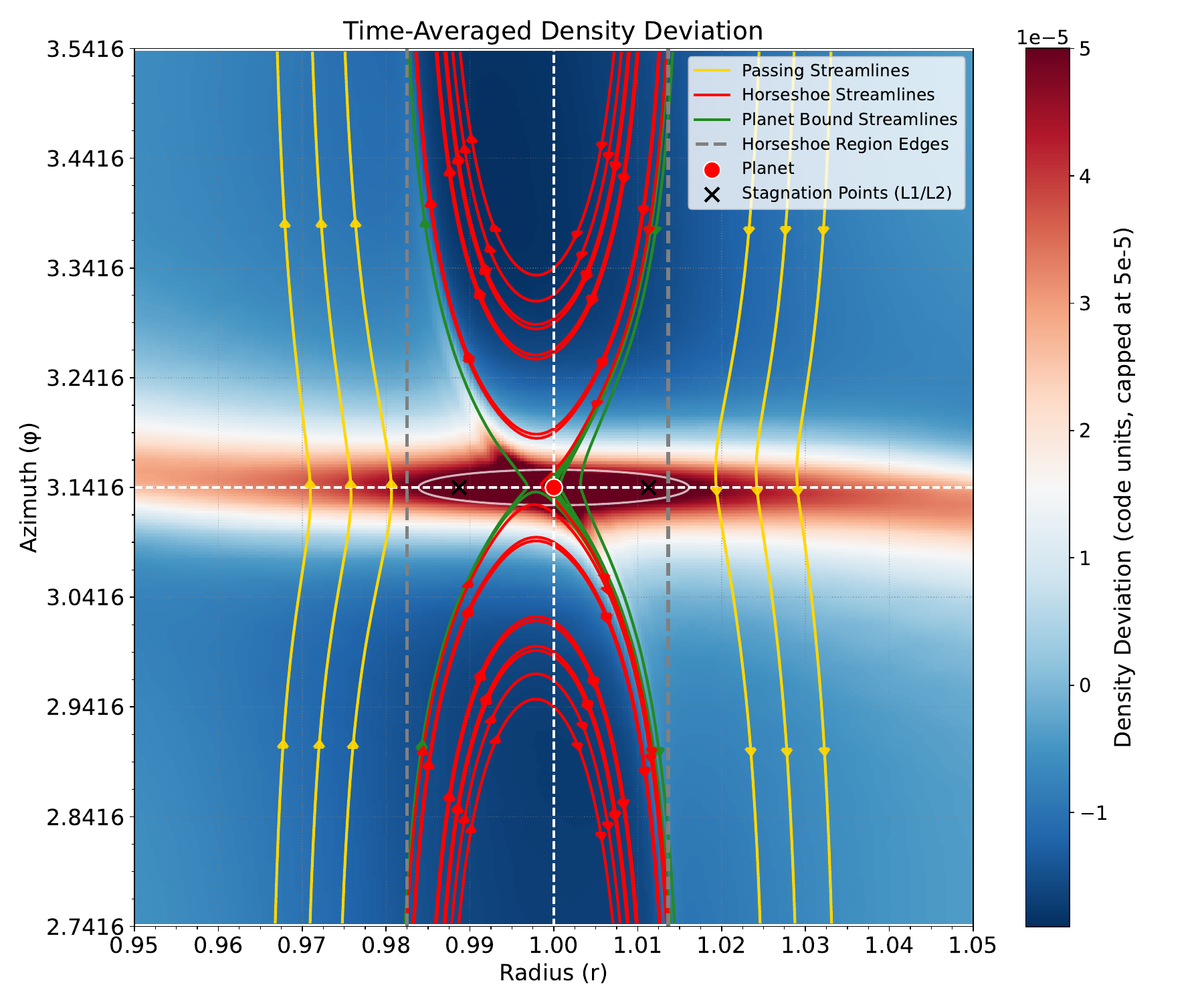}
     \caption{Steady-state fluid streamlines and density field in the vicinity of a low mass planet with $q_p=1.26 \cdot 10^{-5}$ and viscosity  $\nu_p=10^{-7}$. The slight radial inward shift of the gas' horse shoe orbits with respect to the planets location is caused by the radial gas pressure gradient.}
     \label{fig:streamlines}
 \end{figure*}

In this section, we briefly revisit the pioneering theoretical work of \PK, who studied the influence of viscous and thermal diffusion on the Lindblad and corotation torques acting on a low mass planet, where planet migration, as well as any effect from the planet on the disk structure is neglected. 
We use this model to test our simulation method. In addition to thermal diffusion, we consider the effect of optically thin cooling. We find that the PK11-model can describe this well with a small modification.

Figure \ref{fig:streamlines} depicts the time-averaged gas density deviation from equilibrium in the corotating frame of a low-mass planet embedded in a viscous disk ($\nu_p=10^{-7}$). The planet, marked by a red circle, is located at $(r, \phi) = (1.0, \pi)$. Overlaid streamlines illustrate the primary gas flow patterns, which were obtained by integrating multiple fluid trajectories based on the time-averaged velocity field using a fourth-order Runge-Kutta solver: passing outer and inner disk material (yellow), U-turning horseshoe orbits (red), and closely bound circumplanetary material (green). The planet's spiral wake is visible as the high-density feature extending radially inward and outward from its location, while a shallow partial gap is carved in the low-density horseshoe region, extending in both azimuthal directions away from the planet. In addition, the Lagrange points "L1" and "L2"(black crosses), which correspond to stagnation points in the flow, and the planets Hill sphere (gray circle) are indicated.

Several key locations are marked. The grey contour outlines the planet's Hill sphere: a circle of radius $r_H = (M_p/3M_*)^{1/3}r_p$, where $r_p$ denotes the planet's radial location, appearing elliptical due to the plot's aspect ratio. The vertical grey dashed lines define the width of the horseshoe region, determined numerically from the minimum and maximum radial extents of the traced horseshoe streamlines (see Appendix \ref{xs_discussion}).

\subsection{Lindblad torque}\label{sec:lindblad}

We first focus on the Lindblad torque, induced by spiral density waves excited by the planet at adjacent Lindblad resonance locations, and superimposing to form the characteristic two-lobe spiral wake (e.g. \citealt{ogilvie2001}).

According to \PK, the Lindblad torque in the presence of energy dissipation can be written as
\begin{equation}\label{eq:lindblad}
    \Gamma_L =\frac{\Gamma_0}{\gamma_{\mathrm{eff}}} \left(-2.5 -1.7 q + 0.1 s \right).
\end{equation}
with the torque scaling
\begin{equation}\label{eq:scale}
  \Gamma_0 = \left(\frac{q_p}{h_p}\right)^2 \Sigma_p r_p^4 \Omega_p^4  
\end{equation}
where $\Sigma_p$, $h_p$ and $\Omega_p$ denote surface mass density, disk aspect ratio and orbital frequency at the planet radial location, $r_p$, respectively, and with the planet to star mass ratio $q_p = M_p/M_*$.
This is the same expression as in the adiabatic limit \citep[see Eq. \ref{adiabatic_torque_formula} and][]{paardekooper2010}, but with $\gamma$ replaced by an effective $\gameff$. This formula, in principle, now captures the entire cooling regime, ranging from the isothermal limit where $\gamma_{\text{eff}}=1$ to the adiabatic limit where $\gameff = \gamma$.

\PK~ derived $\gameff$ by resorting to linear wave theory and making a suitable approximation. In Appendix \ref{app:gamma_eff_lin}
we repeat this derivation, but for density waves subject to optically thin cooling instead of thermal diffusion, resulting in Eq. (\ref{eq:gammaeff}). There, in addition, we also recover the finding of \citet{miranda2020} that wave damping occurs at intermediate values of cooling.
This damping is expected to affect the planet's Lindblad torque.
Moreover, since vortices excite spiral density waves similar to embedded planets, this damping is also expected to affect the resulting wave -angular momentum transport of the COS. Note, though, that in our 3D simulations below, the considered thermal diffusivity values are significantly smaller than those where damping is expected to be maximal.

\subsection{Corotation torque}\label{sec:corot}
The corotation torque, which we simply refer to as $\Gamma_C$, is more complex than the Lindblad torque. One reason is that linear theory is not suitable for capturing it in its entirety. By $\Gamma_C$ we mean the sum of the linear co-rotation torque and the nonlinear horseshoe drag \citep{paardekooper2009a}.

In simple terms, the corotation torque arises from the conservation of fluid properties as gas executes U-turns relative to the planet. Physically, this torque is a superposition of two distinct conservation mechanisms.

First, in adiabatic inviscid disks, entropy is materially conserved. As co-orbital fluid elements traverse horseshoe orbits, they retain their initial specific entropy. If the disk possesses a radial entropy gradient, fluid moving to a new radius arrives with an entropy distinct from the local ambient material. To maintain pressure equilibrium with the surroundings, this entropy mismatch forces the fluid element to adjust its density, creating a physical non-axisymmetric azimuthal density perturbation that exerts a gravitational torque on the planet.

Second, strictly in barotropic inviscid disks, vortensity (vorticity divided by surface density) is materially conserved. This conservation law constrains the fluid's rotation rate rather than its density. As fluid moves radially, it must adjust its azimuthal velocity (spin up or down) to match its original vortensity \citep{paardekooper2010}. Physically, this represents a direct exchange of angular momentum between the fluid and the planet during the U-turn. While this mechanism manifests as a ``velocity mismatch'' (a change in the rotation profile) rather than a density lump, standard torque formulae \citep[e.g.][]{paardekooper2010} mathematically represent this angular momentum flux as an equivalent surface density perturbation to unify the calculation.

In realistic protoplanetary disks, however, these conservation laws are not absolute. Disks are generally baroclinic rather than barotropic, and processes such as viscous dissipation and thermal diffusion (cooling) act to relax the gradients of vortensity and entropy, respectively. Nevertheless, the mechanisms described above remain the fundamental drivers of the corotation torque. The primary role of viscosity and cooling in this context is to counteract the homogenization of the horseshoe region. By diffusively replenishing the radial gradients that the horseshoe turns attempt to flatten, these non-ideal processes prevent the torque from saturating (vanishing) and allow for a sustained, steady-state torque.
In Appendix \ref{app:corot} we discuss details of the corotation torque in the presence of optically thin cooling, thermal diffusion as well as viscous diffusion, closely following \PK.

\subsection{Total torque formula}\label{sec:total_torque_formula}

The various contributions to the total steady-state torque 
\begin{equation}\label{eq:gamma_tot}
    \Gamma_{tot} = \Gamma_L + \Gamma_C
\end{equation}
discussed in the previous sections can be combined in a comprehensive formula for the total torque of a low mass planet in a non-isothermal, viscous disk, as devised by \PK.
The theoretical corotation torque sums barotropic and entropy-related components:
\begin{equation}\label{eq:GammaC}
\Gamma_C = \Gamma_{C,baro} + \Gamma_{C,ent}.
\end{equation}
Each component blends a non-linear horseshoe drag term ($\Gamma_{HS,x}$) and a linear corotation torque term ($\Gamma_{C,lin,x}$), modulated by functions $F(p_{\nu,\chi})$, $G(p_{\nu,\chi})$, and $K(p_{\nu,\chi})$. These dimensionless functions take values in the interval $[0,1]$ and regulate the torque's saturation level and its transition between linear and non-linear regimes. The barotropic part is (\PK, Eq. 52):
\begin{equation}
\begin{split}
\Gamma_{C,baro} = &\Gamma_{HS,baro} F(p_\nu) G(p_\nu) \\
\quad & + \Gamma_{C,lin,baro} (1 - K(p_\nu)),
\end{split}
\end{equation}
and the entropy-related part is (\PK, Eq. 53):
\begin{equation}
\begin{split}
\Gamma_{C,ent} = & \Gamma_{HS,ent} F(p_\nu)F(p_\chi)\sqrt{G(p_\nu)G(p_\chi)} \\
\quad & +\Gamma_{C,lin,ent} \sqrt{(1-K(p_\nu))(1-K(p_\chi))},
\end{split}
\end{equation}

with
\begin{align}
\xi &= q - (\gamma - 1) s\label{eq:zeta}, \\
\Gamma_{HS,baro} &= \frac{\Gamma_0}{\gamma_{\mathrm{eff}}} 1.1 \left(1.5 - s \right), \\
\Gamma_{C,lin,baro} &= \frac{\Gamma_0}{\gamma_{\mathrm{eff}}} 0.7 \left(1.5 - s \right), \\
\Gamma_{HS, ent} &= \frac{\Gamma_0}{\gamma_{\mathrm{eff}}^2 }7.9 \, \xi, \\
\Gamma_{C,lin, ent} &= \frac{\Gamma_0}{\gamma_{\mathrm{eff}}^2} \left( 2.2 \gameff - 1.4 \right) \xi.\label{eq:linent}
\end{align}

In summary, $F(p_{\nu,\chi})$ accounts for saturation at low diffusion (large $p_{\nu,\chi}$), while the complementary pair $G(p_{\nu,\chi})$ and $K(p_{\nu,\chi})$ handles the cut-off at high diffusion (small $p_{\nu,\chi}$). See Appendix \ref{app:transitionfunctions} for more information on these functions. We note that these functions were optimized to match the numerical experiments of \PK, who used a different code than the one used in this study.

Note also that viscosity ($\nu$, via $p_\nu$) influences \textit{both} corotation torque components. For the barotropic torque, it is the sole agent to prevent saturation. For the entropy-related torque, thermal processes (via $p_\chi$) are primary, but viscosity plays a dual role.
The entropy gradient baroclinically generates vortensity perturbations (via the source term $\propto \nabla S \times \nabla P$, see Eq.~30 in Paardekooper et al. 2010). Viscosity diffuses these secondary vortensity gradients, preventing them from causing premature saturation, hence the $F(p_\nu)$ factor in $\Gamma_{C,ent}$  (\PK).
   
   Furthermore, high viscosity (small $p_\nu$) disrupts the coherent horseshoe flow pattern via momentum exchange between adjacent streamlines, washing out structures faster than the U-turn time. This impacts entropy transport as well, so $G(p_\nu)$ and $K(p_\nu)$ also modulate the entropy term.
Consequently, sufficiently high viscosity ($p_\nu \ll 1$) drives the entire corotation torque towards linearity.

Note that equations (\ref{eq:zeta})---(\ref{eq:linent}) that describe the scaled unsaturated torques are based on a razor thin disk model. However, we can apply them directly to our 3D vertically unstratified simulations, as long as the planet potential does not depend on $z$, and we replace the surface mass density slope $s$ by the volume mass density slope $p$.

  \begin{figure*}
 \centering 
    \includegraphics[width= \textwidth]{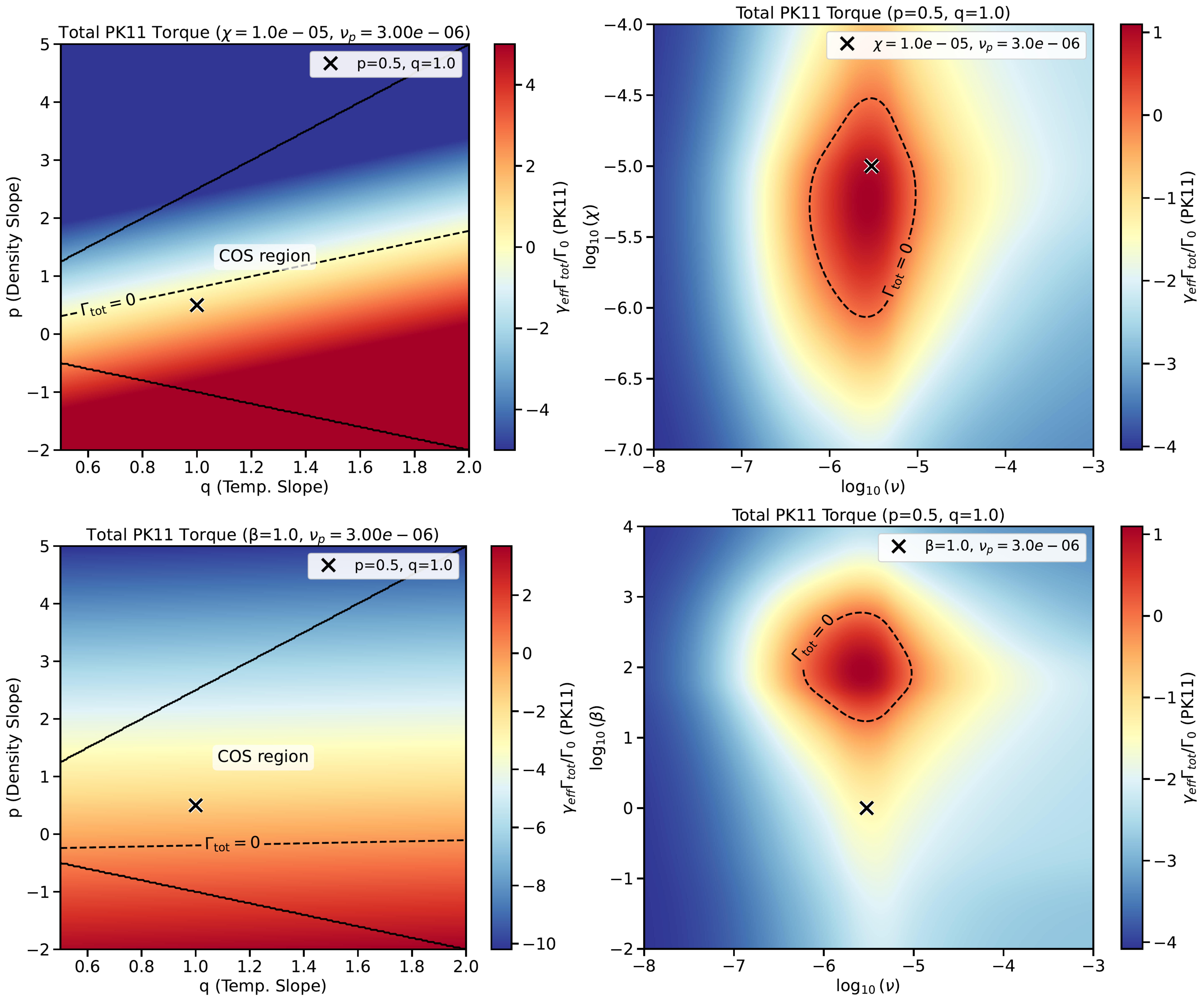}
     \caption{Maps of the total torque resulting from the \PK-model with thermal diffusion (upper panels) and $\beta$-cooling (bottom panels). The used parameters are $h_p=0.1$, $q_p=2.52\cdot 10^{-5}$, $b/h_p=0.4$ and $\gamma=1.4$. The left panels vary the disk structure via $p$ and $q$ at fixed viscous and thermal diffusion, while the right panels vary $\chi_p$ and $\nu_p$ at fixed disk structure. The entire region between the black solid lines yields $N_r^2<0$ via Eq. (\ref{eq:nr22}) in Section \ref{sec:hydro} and is therefore susceptible to the COS.}
     \label{fig:GAM_tot}
 \end{figure*}

In order to apply the above model to simulations using optically thin cooling instead of thermal diffusion, 
we need a mapping of our $\beta$-cooling parameter to an "equivalent" thermal diffusion coefficient $\chi_{equiv}$ used in $p_\chi$. 
A more rigorous approach would be a re-derivation of the transition functions F,G and K suitable for modeling the effect of cooling on the entropy related corotation torque. However, this is beyond the scope of this paper. Moreover, it turns out that a simple mapping results in good enough agreement for our purposes.
We assume the relationship involves an effective length scale $L_{\textrm{eff}}$ over which cooling balances thermal perturbations:
\begin{equation}\label{eq:mapping}
    \chi_{equiv} = \frac{L_{\textrm{eff}}^2}{t_{cool}} = \frac{L_{\textrm{eff}}^2 \Omega_p}{\beta},
\end{equation}
As relevant length scale we choose $L_\text{eff}=x_s$
as it represents the physical size of the region where the linear corotation torque and horseshoe dynamics dominate.

\subsection{Torque maps based on 2D theory}
It is useful to first obtain an overview of theoretical expectations for the total torque predicted by the \PK model in a disk that meets the criteria to trigger COS and, in addition, accounts for viscous diffusion associated with the resulting turbulence.

Evidently, it is a priori not clear whether COS turbulence can be described by a simple laminar Navier-Stokes viscosity in the present context, nor do we know its exact values before running actual simulations.

Figure \ref{fig:GAM_tot} shows the total torque (\ref{eq:gamma_tot}) for thermal diffusion (top panels) and beta cooling (bottom panels). The left panels show the torque for fixed $\nu_p=3 \cdot 10^{-6}$ and $\chi_p = 10^{-5}$ or $\beta=1$, with $h_p=0.1$, respectively, for varying power law slopes $p$ and $q$, whereas the right panels vary the cooling and viscosity, for fixed $p=0.5$ and $q=1$. The left panels also mark the region where the COS can occur based on disk structure alone via Equation (\ref{eq:nr22}). The crosses in each panel indicate the values in the color map used in the corresponding other panel. Note that in a 3D unstratified model, as used for the simulations including COS presented below, the surface density slope $s$ in the theoretical torque formulae (\ref{eq:gamma_tot})---(\ref{eq:linent}) is replaced by the volume density slope $p$. This point will be reiterated in Section \ref{sec:results}.

What these plots show is that for thermal diffusion, a wide variety of disk structures can result in a positive predicted total torque and, in principle, also support COS. This is exactly the regime we want to explore in this study. On the other hand, in the $\beta$-cooling case we find that the cooling times corresponding to the most positive torques, $\beta\sim 100$, are not optimal for COS (which prefers shorter cooling times $\beta\simeq 1$).

\section{Hydrodynamic Model}\label{sec:hydro}
\subsection{Governing Equations}

The gas dynamics are governed by the following equations for mass, momentum, and internal energy, solved in a reference frame co-rotating with the planet at angular velocity $\mathbf{\Omega}_p$:
\begin{align}
    \frac{\partial \rho}{\partial t} + \nabla \cdot (\rho \mathbf{v}) &= 0 \label{eq:cont} \\
    \frac{\partial \mathbf{v}}{\partial t} +  (\mathbf{v} \cdot \nabla) \mathbf{v} &= -\frac{1}{\rho} \left(\nabla P + \nabla \cdot \mathbf{f}\right) - \nabla\Phi_{\text{eff}} \nonumber \\
    & \quad - \nabla\Phi_p - 2(\mathbf{\Omega}_p \times \mathbf{v}) \label{eq:mom} \\
    \frac{\partial \epsilon}{\partial t} + \nabla \cdot (\epsilon \mathbf{v}) &= -P(\nabla \cdot \mathbf{v}) - \Lambda_{\text{cool}} + Q_{\text{diff}}\label{eq:energy}.
\end{align}
In the equations above, $\mathbf{v}$ is the velocity in the rotating frame and $\epsilon=P/(\gamma-1)$ is the internal energy density. The viscous stress tensor, $\mathbf{f}$, is given by
\begin{equation}
    \mathbf{f} = \rho\nu[\nabla \mathbf{v} + (\nabla \mathbf{v})^T - \frac{2}{3}\mathbf{I}(\nabla \cdot \mathbf{v})],
\end{equation}
where $\nu$ is the kinematic viscosity. Stellar gravity is contained within the effective potential, $\Phi_{\text{eff}} = \Phi_* - \frac{1}{2}|\mathbf{\Omega}_p \times \mathbf{r}|^2$. The energy equation includes compressional work alongside cooling and thermal diffusion terms, $\Lambda_{\text{cool}}$ and $Q_{\text{diff}}$, which are specified in Section~\ref{ssec:thermal}.

The system is closed with the ideal gas equation of state:
\begin{equation}
    P = (\gamma-1)\epsilon = \frac{\mathcal{R}}{\mu}\rho T,
\end{equation}
where $T$ is the temperature, $\mathcal{R}$ is the ideal gas constant, and $\mu$ is the mean molecular weight.

\subsection{Disk Equilibrium State}
We assume an initial axisymmetric disk equilibrium where the gas volume mass density and temperature follow radial power-law profiles:
\begin{equation}\label{eq:eq_profiles}
\rho_{eq}(r) = \rho_{p} \left(\frac{r}{r_p}\right)^{-p}, \quad T_{eq}(r) = T_{p} \left(\frac{r}{r_p}\right)^{-q}.
\end{equation}
where $r_p$ is the radial location of the planet\footnote{In simulations without a planet, the mass of the planet is effectively set to zero.} and $\rho_{p}$ and $T_{p}$ are corresponding density and temperature values. The equilibrium azimuthal velocity is given by $v_\varphi = r\Omega(r)$, where the orbital frequency $\Omega(r)$ is the Keplerian frequency corrected for the radial pressure gradient:
\begin{equation}
    \Omega(r) = \Omega_K(r) \left[1 - h^2(p+q)\right]^{1/2},
\end{equation}
where $\Omega_K(r) = \sqrt{GM_*/r^3}$ is the local Keplerian frequency and $h=H/r$ is the disk's aspect ratio. Following \PK, we adopt a radial profile for the viscosity to ensure this initial state is a stationary solution with no viscous accretion, which requires $\nu \propto r^{p-1/2}$.

Using Eq. \ref{eq:nr2}, the buoyancy frequency profile for the above power-law disks is given by 
\begin{equation}\label{eq:nr22}
N_r^2  =  -\frac{1}{\gamma} h^2 \OmK^2 \left(p+q\right)\left(q+\left[1-\gamma\right]p\right). 
\end{equation}
As discussed, a necessary condition for the COS is $N_r^2 < 0$. For typical temperature gradients ($q>0$), this requirement translates to flat or rising density profiles \citep{lesur2023}. The COS also requires finite thermal losses, described next.

\subsection{Thermal Physics}\label{ssec:thermal}

A key component of this study is the implementation of different thermal physics prescriptions to investigate their effect on the COS and its interaction with the planet. We consider both a simple cooling law and three distinct models for thermal diffusion.

\subsubsection{Optically-thin Cooling}

To model thermal relaxation in optically thin regions of the disk, we employ a ``beta cooling'' prescription. The cooling term, $\Lambda_{\text{cool}}$, drives the gas temperature $T$ back towards a prescribed profile $T_{eq}$ over a characteristic cooling timescale $t_c$:
\begin{equation}
\Lambda_{\text{cool}} = \frac{P - \rho T_{eq}}{t_c} = (\gamma-1) \frac{\mathcal{R}}{\mu}\frac{\rho(T - T_{eq})}{t_c}. 
\end{equation}
The cooling time is parameterized by the constant dimensionless cooling parameter: \citep{klahr2014}:
\begin{equation}
    \beta = t_c(r) \Omega_K(r).
\end{equation}
Such that $\beta$ is the cooling time in code units at the planet's location. 
We take $T_{eq}$ to be the initial equilibrium profile described by Eq. (\ref{eq:eq_profiles})

\subsubsection{Thermal Diffusion Models}\label{sec:thermal}

We investigate three different mathematical formulations for the thermal diffusion term, $Q_{\text{diff}}$, models A,B and C as described below, which were chosen to explore different physical and numerical approximations for heat transport and relaxation. This is also done to compare our results to those of \PK, who considered models A and B.

\paragraph{Model A: Laplacian on Entropy}
This model diffuses the specific entropy 
and is formulated to smooth out entropy gradients:
\begin{equation}
    Q_{\text{diff}} = (\gamma-1) \chi \rho T \nabla^2 S 
\end{equation}
where $\chi$ is the thermal diffusion coefficient or thermal diffusivity. Note that for power-law disks, $\nabla^2 S = 0$ in cylindrical coordinates. Hence $Q_{\text{diff}}=0$ and the disk remains in thermal equilibrium initially.

\paragraph{Model B: Heat Conduction}
This model implements standard heat conduction based on Fourier's law, representing the physical transport of heat down a temperature gradient:
\begin{equation}
    Q_{\text{diff}} = (\gamma-1) \nabla\cdot(\chi_T(r) \nabla T)
\end{equation}
This formulation directly models thermal energy flux and has the advantage of affecting the temperature field directly. In this case, the thermal diffusivity is required to follow a power law $\chi_T(r) \propto r^{q}$ for the disk to be in initial thermal equilibrium.

\paragraph{Model C: Temperature Relaxation via Laplacian}
The third model is a numerical prescription designed to relax temperature fluctuations, $\delta T = T - T_{eq}$, back to the equilibrium profile. The term is:
\begin{equation}
    Q_{\text{diff}} = (\gamma-1) \chi \nabla^2(T-T_{eq})
\end{equation}
It acts as a diffusive relaxation mechanism where the rate of relaxation is sensitive to the spatial scale of the temperature perturbations. In this work we use a fixed value of $\gamma=1.4$.

\subsection{Planetary potential}\label{ssec:potential}

We follow \citet{paardekooper2010} and assume the gravitational potential of the planet, $\Phi_p$, is modeled as a softened point mass to avoid a singularity at the planet's location, $(r_p, \phi_p)$. This softening is also essential for approximating 3D vertical averaging effects in 2D razor-thin simulations. The potential is given by :
\begin{equation}
\Phi_p(r, \varphi) = -\frac{GM_p}{\sqrt{r^2 + r_p^2 - 2rr_p\cos(\varphi-\varphi_p) + b^2 r_p^2}}    
\end{equation}
where $M_p$ is the planet mass and $b$ is the dimensionless softening length. Consistent with our vertically unstratified disk model, the potential has no vertical dependence; it is evaluated at the midplane ($z=0$) and applied uniformly at all heights in the 3D simulations. The softening length is typically scaled with the local disk pressure scale height at the planet's location, $H_p$, and throughout this work we use $b = 0.4 h_p$ \citep{paardekooper2010}. Note that in our code's units $r_p=1$ and $\varphi_p=0$ are constant, as we consider a non-migrating planet.

\subsection{Simulation setup}\label{sec:numerics}

We conduct hydrodynamic simulations to solve equations (\ref{eq:cont})---(\ref{eq:energy}) using FARGO3D\footnote{\url{http://fargo.in2p3.fr}} \citep{fargo3d, llambay2019}. Boundary conditions are periodic in both azimuthal and vertical directions for all quantities. 
The radial boundary conditions extrapolate the equilibrium values of mass density and velocity into the ghost zones. In simulations containing a planet, damping zones are applied at the radial boundaries to restore all field variables to their equilibrium values and to mitigate spiral wave reflection. In 3D simulations without a planet, damping boundaries are generally not applied, since this could potentially mitigate, or at least substantially slow down, the growth of COS turbulence \citep{lehmann2024}. We did, however, run 3D simulations without a planet, including damping zones for select parameter values, to confirm that the COS eventually develops similarly to simulations without damping zones, although it takes several thousand orbits longer to reach the fully saturated state.

Simulations are performed in cylindrical coordinates as defined in Section~\ref{sec:hydro}. We adopt a system of units in which $r_{p} = M_{*} = G = 1$, and set the disk aspect ratio to $h_{p} = H_{p}/r_{p} = 0.05$ in all runs presented in Section \ref{sec:pk11} and $h_0=0.1$ in all runs presented in Section \ref{sec:results}.
All runs presented in Section \ref{sec:results} adopt a computational domain $0.5 \leq r \leq 1.6$ and $-0.25 H_{p} \leq z \leq +0.25 H_{p}$, with the full azimuthal range $0 \leq \varphi \leq 2\pi$ covered in the 3D simulations.
The grid resolution is $N_{r} \times N_{z} \times N_{\varphi} = 1600 \times 75 \times 628$, corresponding to radial and vertical resolutions of approximately $150/H_{p}$ for $h_p=0.1$. As was also done by \citet{lehmann2024} and \citet{lehmann2025}, the azimuthal resolution, about $10/H_{p}$, is significantly lower.
Some simulations presented in Section \ref{sec:pk11} adopt a slightly different domain and resolution, which will be explicitly mentioned. This is done to facilitate a direct comparison with \PK~.

All 3D simulations are executed on GPU nodes equipped with eight NVIDIA~A100 or V100 GPUs. Most runs cover 1,000 reference orbits, with select cases extended further as noted above.

\subsection{Diagnostics}\label{sec:diagnostics}

\subsubsection{Turbulence properties}

As in \citep{lehmann2024}, we characterize the radial turbulent angular momentum transport in our 3D simulations through the dimensionless parameter
\begin{equation}\label{eq:alpha}
\alpha_{r}(t) = 
\frac{\langle \rho v_{r} \, \delta v_{\varphi} \rangle_{r\varphi z}}
     {\langle P \rangle_{r\varphi z}},
\end{equation}
where the brackets denote averages over the indicated spatial dimensions, and 
$\delta v_{\varphi}$ is the deviation of the azimuthal velocity from its equilibrium value.

To further quantify the strength of COS-induced motions, we compute the root-mean-square (RMS) velocity as
\begin{equation}
\mathrm{RMS}(v_i) = \sqrt{\langle v_i^2 \rangle_{r\varphi z}}.
\end{equation}
where $i$ is either $r$, $\varphi$ or $z$.
Following \citet{lehmann2025}, radial averages are taken over $0.9 < r/r_p < 1.2$ to minimize boundary effects. 
This asymmetric interval is adopted because the inner boundary region typically exhibits enhanced hydrodynamic activity. 
Azimuthal averaging is performed over the full $2\pi$ domain, and vertical averaging over the entire computational domain.

\subsubsection{Disk-Planet torque}\label{sec:planet_torque}

The gravitational torque exerted by the disk on the planet is computed from the planet potential $\Phi_p$ as
\begin{equation}\label{eq:planet_torque}
\begin{aligned}
\Gamma &= \int_V \rho(\mathbf{r}) \left( \mathbf{r} \times \nabla \Phi_p \right)_z \, dV \\[4pt]
       &= \int \rho(r,\varphi,z)\, r\, \frac{\partial \Phi_p}{\partial \varphi}\, dr\, d\varphi\, dz,
\end{aligned}
\end{equation}
where $(r,\varphi,z)$ denote cylindrical coordinates and the integration extends over the entire simulation domain.
Positive values of $\Gamma$ correspond to angular momentum transfer from the disk to the planet.

  \begin{figure}
 \centering 
 	\includegraphics[width=0.5 \textwidth]{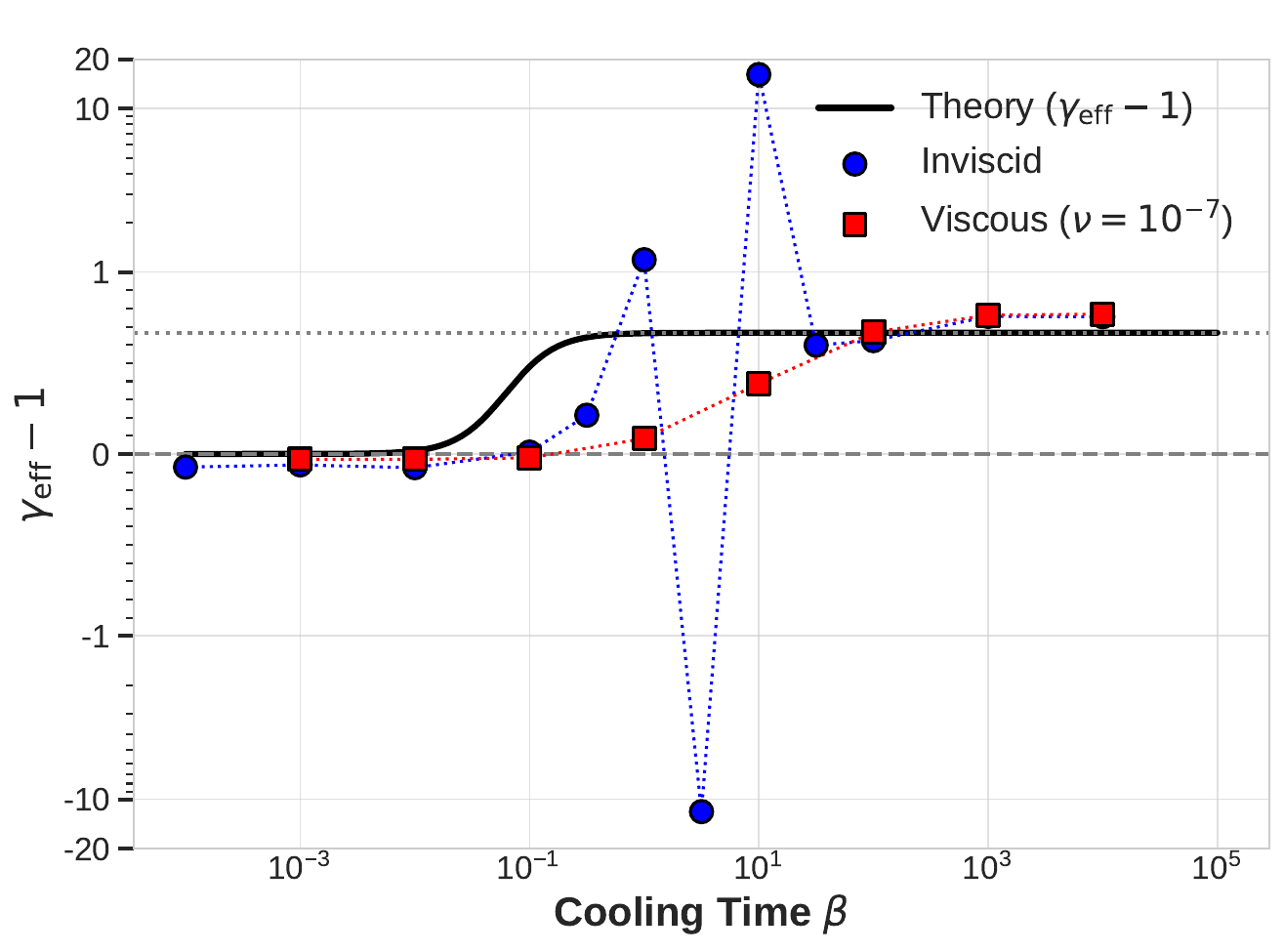}
     \caption{ Measurements of $\gameff$ from the Lindblad torque in 2D simulations with viscosity $\nu=10^{-7}$ (red squares) and inviscid simulations (blue circles) and optically thin cooling indicated on the horizontal axis, as explained in the text.}
     \label{fig:gamma_eff}
 \end{figure}

\section{Results of 2D test simulations }\label{sec:results2D}

For comparison with \PK, we ran a series of 2D simulations, where we replace the volume density slope $p$ in (\ref{eq:lindblad}) by the surface mass density slope $s$. 

\subsection{Lindblad Torque}\label{sec:lindblad_results}

To measure the Lindblad torque, we ran simulations that adopted different cooling times $\beta$ and fixed slopes $s=1.5$, $q=1$, and $\gamma=5/3$, which cover the relevant values of $Q_\beta$ ranging from the isothermal limit to the adiabatic regime. Note that for these parameters, the radial entropy and vortensity gradients, and therefore the corotation torque identically vanishes (see \S\ref{sec:corot}), allowing for a more precise measurement of the Lindblad torque. 

The results are summarized in Figure \ref{fig:gamma_eff} for inviscid ($\nu_p=0$) simulations (red squares) and simulations with $\nu_p=10^{-7}$ (blue circles). Here, we measure the torque on the planet and use Eq. (\ref{eq:lindblad}) to infer the corresponding $\gamma_\mathrm{eff}$, then compare it with the theoretical value in Eq. (\ref{eq:gammaeff}).

Our viscous results are similar to the results of \PK, i.e. we find good agreement in the isothermal and adiabatic limits, whereas the transition region is not well captured by the simple analytic estimate Eq. (\ref{eq:gammaeff}).
It is worth noting that the measured values of $\gamma_{\mathrm{eff}}$ exhibit a notable deviation from the theoretical prediction in the intermediate cooling regime, $\beta \sim 0.1 - 10$. It is not fully clear if this discrepancy arises from the linear wave damping discussed Appendix \ref{app:gamma_eff_lin}, which reaches its maximum efficiency in exactly this parameter range. The analytic torque formula (Eq. \ref{eq:lindblad}) is adapted from the adiabatic limit by modifying the gas stiffness ($\gamma \to \gamma_{\mathrm{eff}}$); 
\begin{figure*}
\centering
\includegraphics[width=\textwidth]{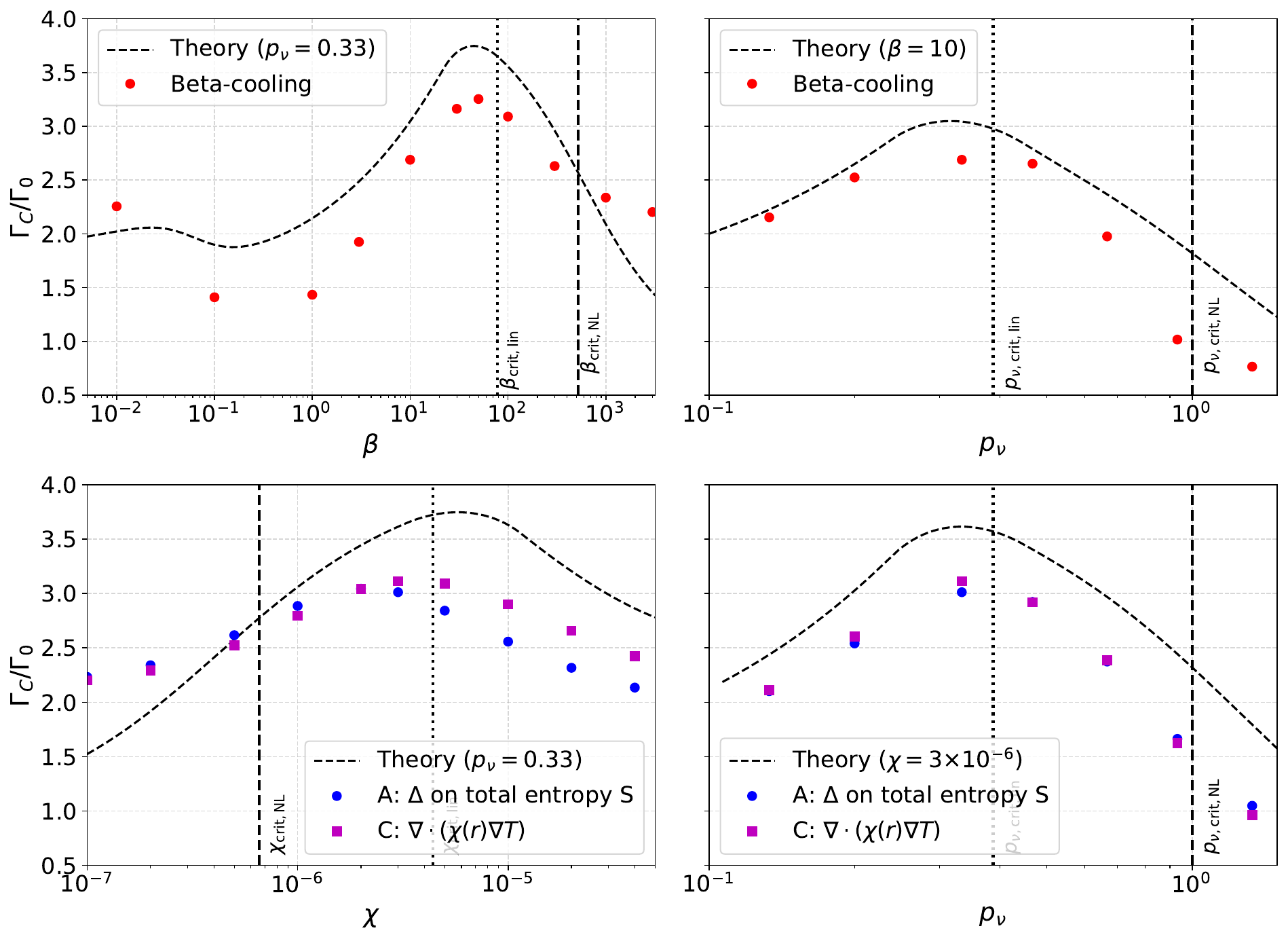} 
\caption{Measurements of the scaled corotation torque $\Gamma_C/\Gamma_0$ in viscous 2D simulations with $q_p=1.26\times 10^{-5}$ and $h_p=0.05$. Symbols indicate simulation results obtained by time-averaging the total disk torque over the final 100 orbits and subtracting the theoretical Lindblad torque ($\Gamma_L$, Eq. \ref{eq:lindblad}). \emph{Upper panels} use $\beta$-cooling, varying $\beta$ at fixed $p_{\nu}=0.33$ (left) and varying $p_{\nu}$ at fixed $\beta=10$ (right). \emph{Lower panels} consider thermal diffusion of varying strength at fixed $p_{\nu}=0.33$ (left) and varying $p_{\nu}$ at fixed $\chi = 3 \times 10^{-6}$ (right). Pink squares denote temperature diffusion (method C) and blue circles entropy diffusion (method A). Theoretical curves (dashed black) are based on the torque formulas (\ref{eq:GammaC})--(\ref{eq:linent}). Vertical dashed and dotted lines indicate critical values for non-linear saturation and the linear transition, respectively. These correspond to critical cooling times (Eqs. \ref{eq:beta_crit_nl}, \ref{eq:beta_crit_lin}) and thermal diffusivities (Eqs. \ref{eq:chi_crit_nl}, \ref{eq:chi_crit_lin}) in the left panels, and critical viscosities (Eqs. \ref{eq:nucrit_final}, \ref{eq:nu_crit_lin}) in the right panels. To facilitate direct comparison, the resolution for the thermal diffusion runs (lower panels) matches the grid of \PK while the $\beta$-cooling runs (upper panels) utilize the same grid as our 3D simulations (see Sect. \ref{sec:results}).}
\label{fig:combined_torque}
\end{figure*}
essentially accounting for the change in the wave's phase velocity. It does not, however, explicitly account for the spatial decay of the wave amplitude due to thermal relaxation. Similarly, the formula neglects viscous damping, which also contributes to wave attenuation but is not captured by the effective sound speed approximation. If these damping lengths become comparable to the width of the torque excitation region, the assumption of freely propagating waves breaks down, potentially leading to the observed deviations.

The inviscid results are identical to the viscous results in the isothermal and adiabatic regime, but show strong departures for intermediate cooling times $\beta=1-10$. The large values of $\gameff$ for these cooling times reflect small time-averaged values of the Lindblad torque and result from the onset of Rossby-wave instability due to sharp potential vorticity features emerging at intermediate cooling time. However, as this behavior is not central to the present study, we do not discuss it further here.

An analogue of Figure \ref{fig:gamma_eff} was presented by \PK~ for the case of thermal diffusion. Our attempt to reproduce this in \textsc{fargo3d} was unsuccessful because the explicit thermal diffusion implementation led to numerical instabilities. In particular, we observe the onset of noise for $\chi_p \gtrsim 10^{-3}$ and simulations crash for $\chi_p \gtrsim 5 \times 10^{-3}$. However, the values used in our physical simulations presented in subsequent sections ($\chi_p \leq 4 \times 10^{-5}$) lie safely outside this problematic regime. For these values, the analogous  parameter to $Q_{\chi}$, as defined by (\ref{eq:qchi}), remains below $0.2$. The values of $\gameff$ for $\chi_p \leq 5 \times 10^{-3}$ drop within the range $~1.67-1.1$, consistent with the theoretical expectation.

\subsection{Corotation Torque}

Figure \ref{fig:combined_torque} presents measurements of the scaled corotation torque, $\Gamma_C/\Gamma_0$, obtained from 2D laminar viscous simulations employing $\beta$-cooling (upper panels) and thermal diffusion (lower panels). The corotation torque is inferred by measuring the total torque exerted on the planet, time-averaged over the final 100 orbits of the simulation (out of a total of 600), and subtracting the theoretical Lindblad torque calculated via Eqs. (\ref{eq:lindblad}) and (\ref{eq:scale}).

The left panels illustrate the dependence on cooling strength at a fixed viscosity of $p_\nu=0.33$ (corresponding to $\nu_p=2.63 \times 10^{-6}$), while the right panels display the variation with viscosity at a fixed cooling parameter. For the thermal diffusion simulations, we utilized models A and C as described in Section \ref{sec:thermal}. We observe small deviations between the different thermal diffusion methods at larger diffusivities, specifically for $\chi \gtrsim 3 \cdot 10^{-6}$.

To ensure consistency across our results, the $\beta$-cooling simulations utilized the same radial domain $r=[0.5, 1.6]$ and grid resolution ($N_r \times N_{\phi} = 3200 \times 628$) as the 3D simulations presented in Section \ref{sec:results}. In contrast, for the thermal diffusion simulations, we adopted the extended domain $r=[0.4,1.6]$ and specific resolution of \PK ($N_r \times N_{\phi} = 2512 \times 924$). This choice was made to maximize the precision of our validation against their benchmark, particularly given that the transition functions in their torque model were calibrated to simulations performed with the \textsc{rodeo} code \citep{paardekooper2006}. We confirmed, however, that the results are robust to these choices. While the specific radial extent makes only a minimal difference for the 2D torque measurements, it becomes important in the 3D simulations where the COS is active near the inner domain boundary.

We note that \PK  reported deviations between different codes on the order of 10--20 percent, depending on the parameter regime. In this context, the general agreement between our thermal diffusion simulations and the theoretical curves reinforces the credibility of our implementation in \textsc{fargo3d}. Moreover, the match is similarly good, if not superior, for the $\beta$-cooling simulations, demonstrating that the assumed mapping (\ref{eq:mapping}) is sufficiently accurate for our purposes.

\section{Planet-disk interaction under the COS}\label{sec:results}

We now examine disk-planet torques in 3D. In this case, the COS can develop, provided that the radial entropy gradient is negative, as also needed to produce a related positive horseshoe drag.

\subsection{COS with thermal diffusion}

\begin{figure}
\centering
\includegraphics[width=0.5 \textwidth]{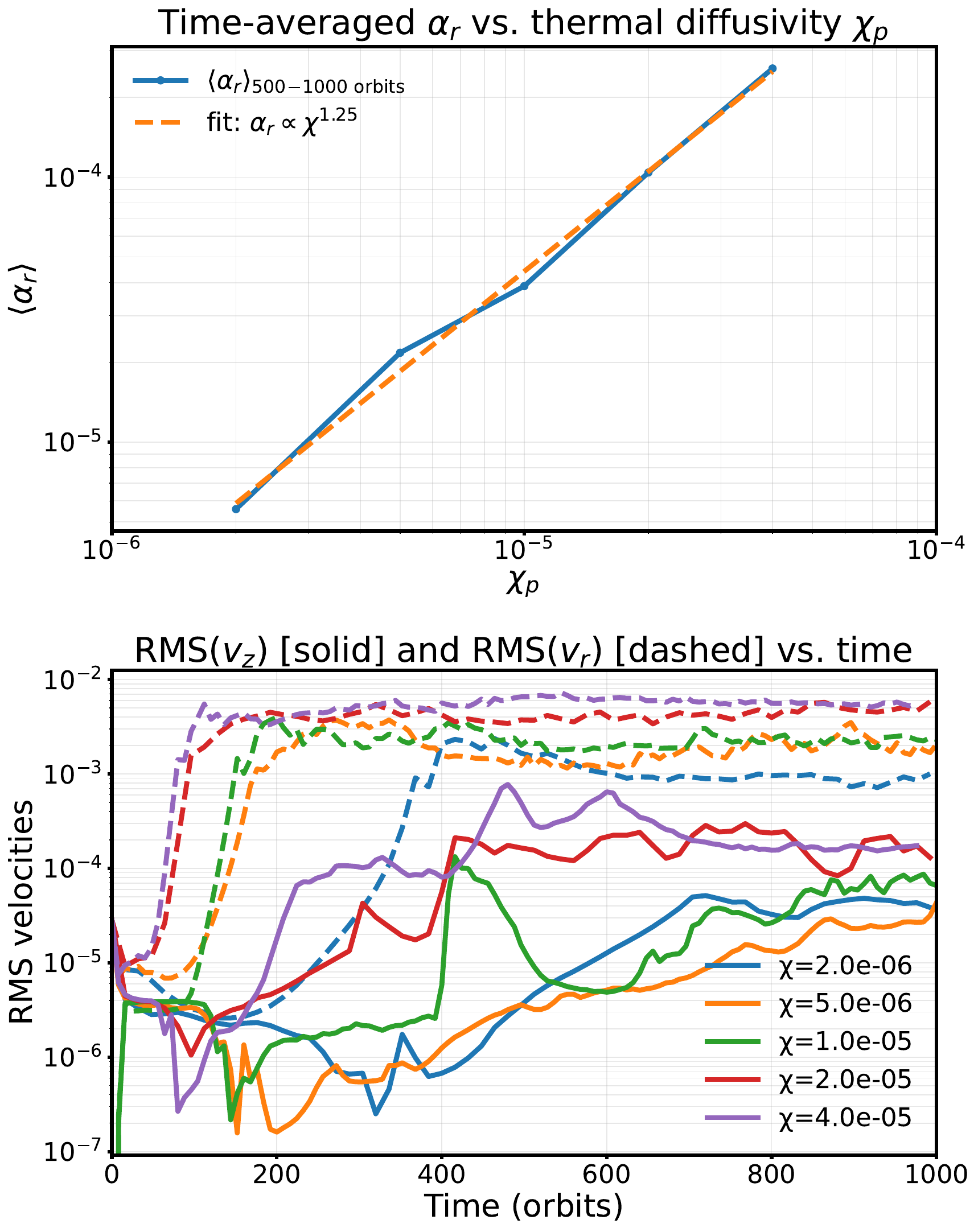}
\caption{Saturation of COS in a disk with thermal diffusion (method $A$). The upper panel shows the measured turbulent radial angular momentum flux as function of thermal diffusivity at the planet's location. The lower panel shows the evolution of RMS radial and vertical velocities as the COS saturates over several hundreds of orbits in the same simulations with different $\chi_p$.}
\label{fig:COS_sat_tdiff}
\end{figure}

\begin{figure}
\centering
\includegraphics[width=0.5 \textwidth]{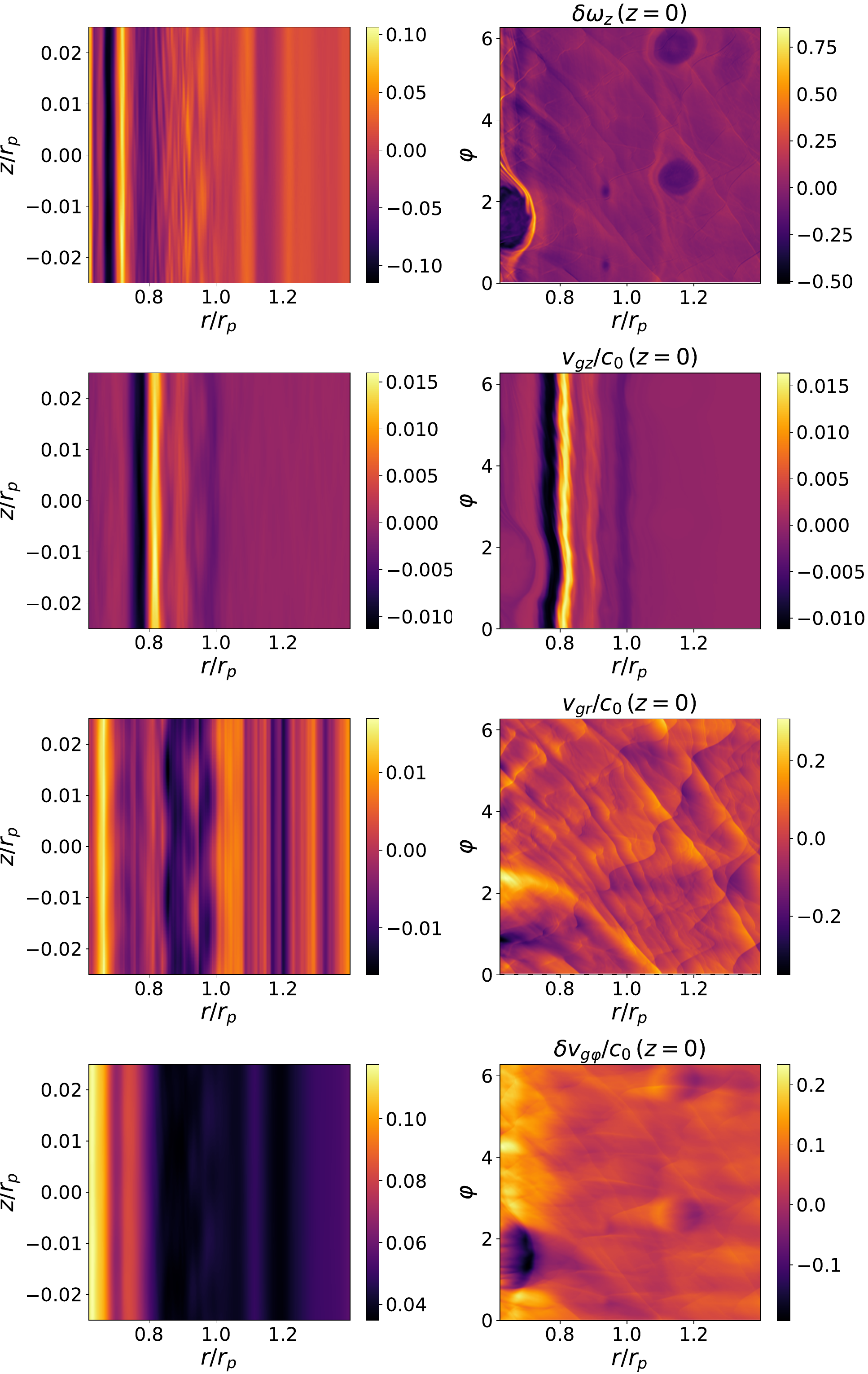}
\caption{Saturation of COS with thermal diffusion method $A$ and $\chi_p=4\cdot 10^{-5}$ around 1000 orbits. From top to bottom the panels show azimuthally averaged meridional views (left) and planar cuts at the mid-plane (right) of the $z$-component of vorticity $\nabla \times \boldsymbol{v}$ in the inertial frame, as well as vertical, radial and azimuthal velocity. Several vortices are present in the domain, most clearly visible in tha azimuthal velocity deviation.}
\label{fig:COS_sat_tdiff_2}
\end{figure}

In this section, we briefly describe the evolution of the COS in our 3D simulations with thermal diffusion, but no planet. Figure \ref{fig:COS_sat_tdiff} shows time-averaged $\alpha_r$ values over the last 500 orbits of simulations using the entropy implementation $A$ of thermal diffusion, with increasing values of $\chi_p$ (top panel), as well as the time evolution of RMS radial (dashed) and vertical (solid) velocities for the same simulations. The dashed curve in the upper panel is a power-law fit to the simulation data, showing a relation
\begin{equation}
    \alpha_r(r) \sim \chi_p^{1.25}.
\end{equation}
The radial angular momentum transport, quantified by $\alpha_r$, is mainly due to vortices excited by the COS, and which excite spiral density waves. This is illustrated in Figure \ref{fig:COS_sat_tdiff_2} for the simulation with $\chi_p=4\cdot 10^{-5}$.
The level of turbulence found here is somewhat smaller than what was  reported by \citet{lehmann2024}. The main reason is that they considered optically-thin cooling disks with a steeper temperature slope $q=2$, resulting in significantly more vigorous COS. We also ran simulations with optically-thin cooling using $\beta=1-10$, and find a milder saturation amplitudes of the COS for the current disk structure than with thermal diffusion.

We also note that simulations employing methods $B$ nd $C$ for thermal diffusion do not exhibit a similar monotonic behavior of $\alpha_r$ as a function of $\chi_p$. Instead, $\alpha_r$ peaks  around $\chi_p\sim 5\cdot 10^{-6}$ and decays thereafter. We will briefly come back to this point in Section \ref{sec:caveats}.

\subsection{Effect of COS on disk-planet torque}\label{sec:cos_torque}

Now that we have tested our simulation setup against the theory of \PK~ and briefly examined the evolution of the COS enforced by thermal diffusion, we can step forward and add a new element, turbulence generated by the COS.
We generally find that the torques resulting in these simulations are highly oscillatory, showing stochastic variations, a feature also reported by \citet{stollpicogna2017} for the VSI. 

\begin{figure*}
\centering
\includegraphics[width=0.9 \textwidth]{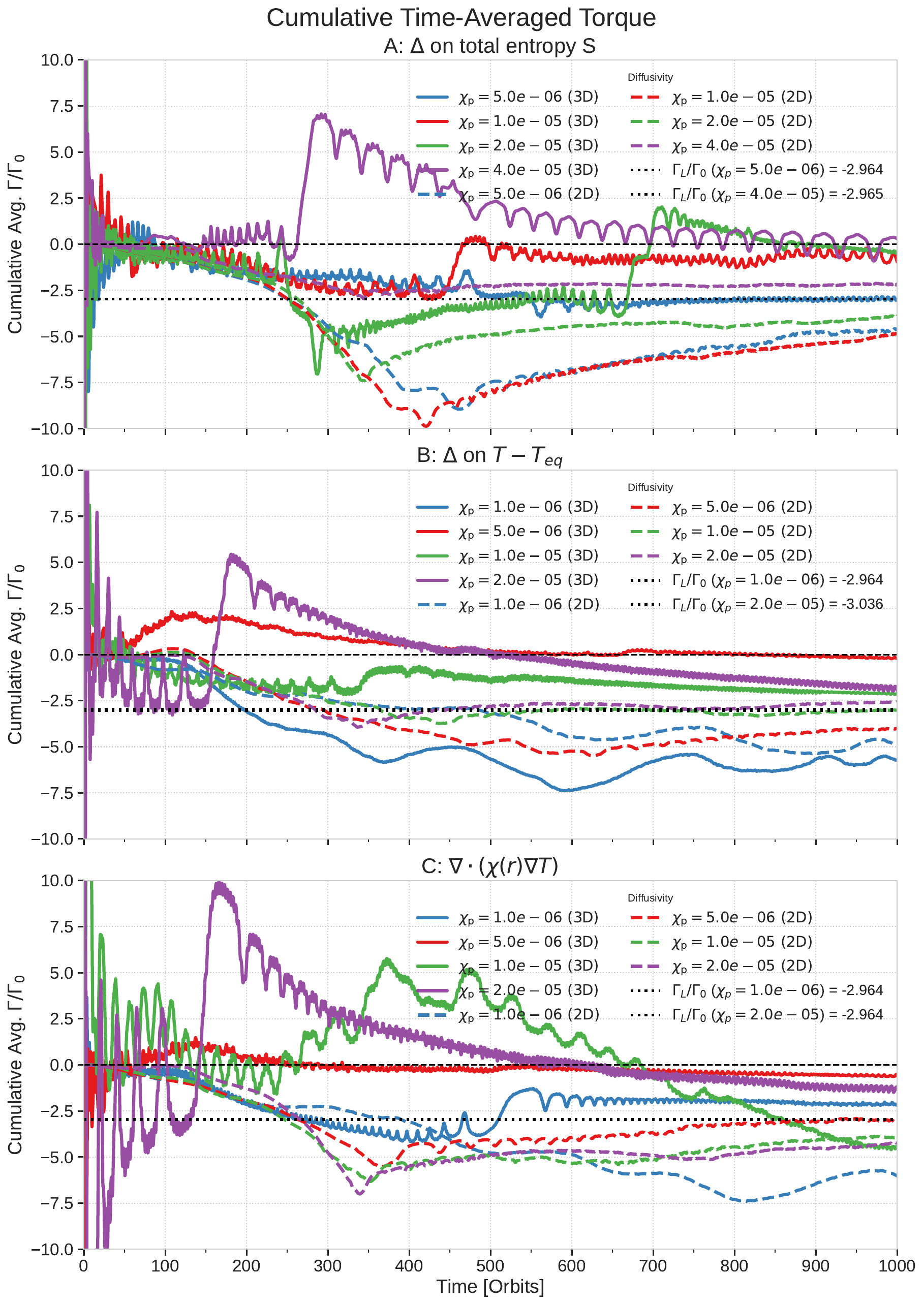}
\caption{Measured cumulative time-averaged disk-planet torque in 2D and 3D simulations adopting different values of thermal diffusion strength $\chi_p$. The black dotted horizontal line marks the (narrow) range of (inviscid) Lindblad torques corresponding to the different simulations.}
\label{fig:torque_cumu}
\end{figure*}

\begin{figure*}
\centering
\includegraphics[width=0.9 \textwidth]{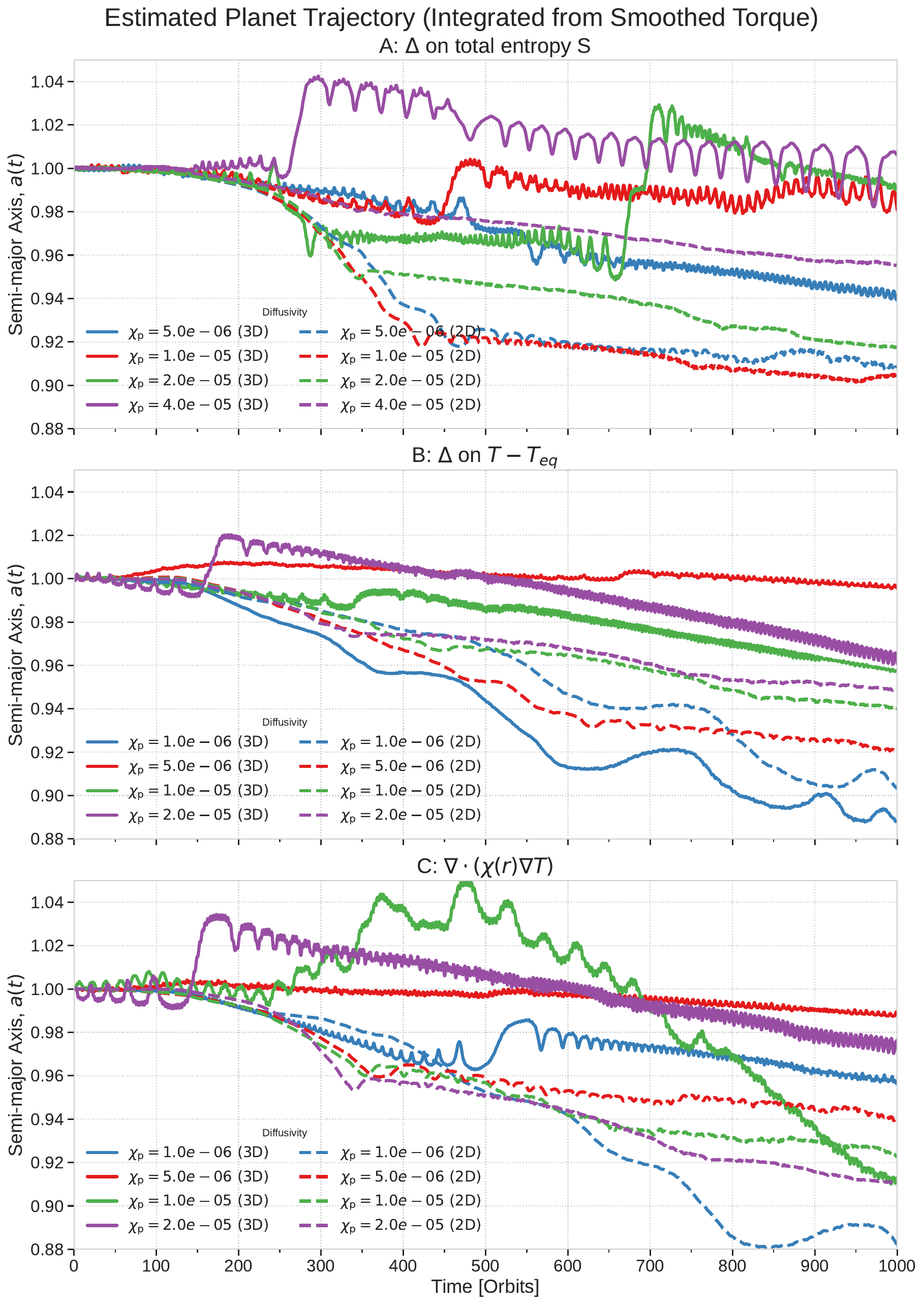} 
\caption{Same as figure \ref{fig:torque_cumu}, but now showing the predicted planet's semi-major axis in code units following from Eq. (\ref{eq:ode}) based on the measured disk-planet torque.}
\label{fig:torque_traj}
\end{figure*}

Note that for our 3D vertically unstratified simulations the torque scaling factor (\ref{eq:scale}) is given by 
$\Gamma_0 = \left(q_p/h_p\right)^2 \rho_p \, \Delta z \, r_p^4 \Omega_p$
which depends on the vertical size of the simulation domain.
Compared to a razor-thin disk model (as assumed in Section \ref{sec:pk11}), we simply replace $ \Sigma_p \to \rho_p \Delta z $ and $s \to p$
in Eq. (\ref{eq:scale}) as well as
the Lindblad and corotation torque expressions (\ref{eq:lindblad}) and (\ref{eq:zeta})---(\ref{eq:linent}).

Figure \ref{fig:torque_cumu} presents the cumulative time-averaged\footnote{The raw torques are highly oscillatory and presenting them does not provide much insight.} disk-planet torque for different implementations of thermal diffusion (the different panels), as well as different values of the thermal diffusivity $\chi_p$. Each panel compares inviscid 3D (solid) and 2D radial-azimuthal (dashed) simulations. The only difference between 2D and 3D simulations is the development of COS in 3D simulations, since our disk is vertically unstratified. These results clearly demonstrate the effect of COS on the torque exerted on the planet. Notably, only 3D simulations with the COS yield positive cumulative torques over long timescales.

The effect is even more clear if we attempt to compute prospective planet migration trajectories $a(t)$.
We do this by numerically integrating the torque exerted by the disk. The torque, $\Gamma$, is related to the rate of change of the planet's orbital angular momentum, $L$, by
\begin{equation}
    \Gamma = \frac{dL}{dt}
\end{equation}
For a planet of mass $m_p$ on a circular orbit around a star of mass $M_*$, the angular momentum is $L = m_p \sqrt{G M_* a}$. Differentiating this with respect to time and equating it to the torque gives the rate of change of the semi-major axis, which can be re-arranged to yield
:
\begin{equation}\label{eq:ode}
    \frac{d\sqrt{a}}{dt} = \frac{\hat{\Gamma}(t)}{\sqrt{G M_*}}
\end{equation}
with the specific torque $\hat{\Gamma} \equiv \Gamma / m_p$, which is numerically more stable than solving directly for $a$.

The torque from the 3D simulations, which develop COS, is highly oscillatory, and undergoes stochastic variations. To ensure a stable integration that captures the net migration trend, we first apply a boxcar average to the specific torque time series, $\hat{\Gamma}(t)$, smoothing it over a timescale of 1 orbit\footnote{The torque is sampled every 0.05 orbits during the simulations. However, We do not find any significant impact of this pre-smoothing on the trajectories}. This yields a net specific torque that drives long-term migration. We then solve the differential equation (\ref{eq:ode})
using a 4th-order Runge-Kutta integrator, with linear interpolation used to find the torque at the intermediate time steps required by the algorithm. The final trajectories are then given by $a(t)$, and are shown in Figure \ref{fig:torque_traj}.

Note that since the computed trajectories depend on the physical (unscaled) torque $\Gamma$, we need to scale the 2D and 3D simulations to the same initial surface mass density profile so as to achieve the same torque scaling $\Gamma_0$ for direct comparison. Here we scale our 3D simulations to match our 2D simulations. This is required because we initialize our 2D simulations with an input value of $\Sigma_p$, whereas we initialize our 3D simulations with a $z$-independent volume density $\rho_p = \frac{\Sigma_p}{\sqrt{2 \pi} H_p}$, so that in the latter the surface mass density, and therefore the torque scale $\Gamma_0$, depends on the chosen vertical domain size.

The results shown in Figure~\ref{fig:torque_traj} suggest that small vortices and other non-axisymmetric structures introduce a form of "base noise" that elevates the torque in 3D simulations relative to the inviscid 2D cases.   More notably, the distinct upward "kicks" in the $\Gamma(t)$ curves observed in many 3D simulations are the most striking feature distinguishing the two sets of results. These torque spikes arise from large vortices that migrate inward—crossing the planet’s orbit from outside to inside—and transfer angular momentum during their horseshoe turn.

\begin{figure}
 \centering 
 	\includegraphics[width= 0.5 \textwidth]{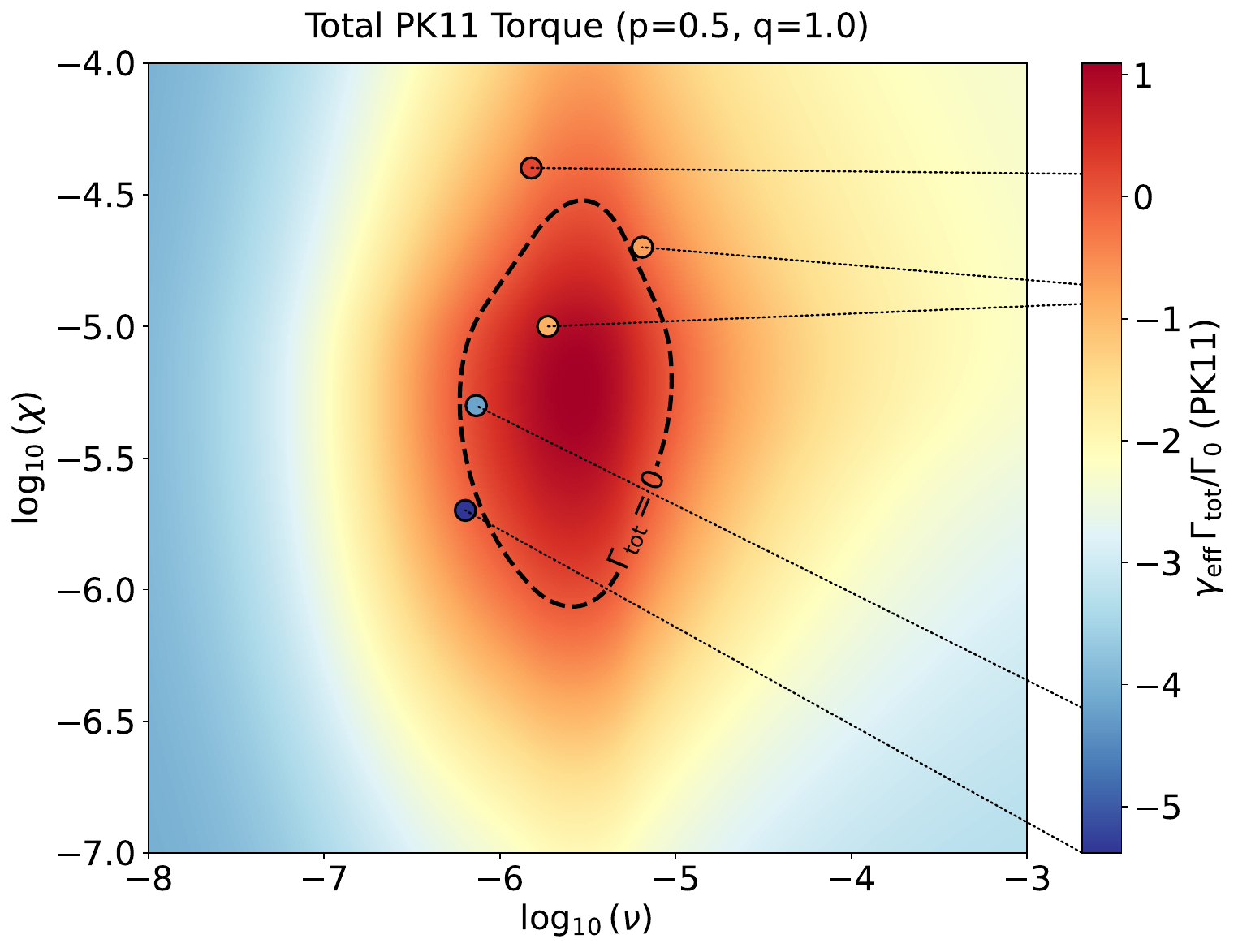}
     \caption{The color map is the same as in Figure \ref{fig:GAM_tot}. The circles represent the time-averaged (over the last 800 orbits) torque and turbulent viscosity (horizontal axis) measured in 3D simulations employing method $A$ for thermal diffusion $\chi_p$ indicated by the vertical axis.}
     \label{fig:torque_map}
 \end{figure}

Overall, the torques measured in our 3D simulations remain significantly less positive than the theoretical values predicted when using the turbulent viscosity inferred from the simulations themselves. This behavior is shown in Figure~\ref{fig:torque_map} for simulations employing thermal diffusion prescription~A across the range $\chi_p = (2\text{–}40)\times 10^{-6}$. The case with $\chi_p = 4\times 10^{-5}$ is a notable exception: in this run, the torque \emph{exceeds} the theoretical expectation.

The turbulent viscosity measured in these simulations is somewhat larger than in the corresponding precursor runs without a planet (Figure~\ref{fig:COS_sat_tdiff}), likely due to the planet triggering the formation of additional vortices by the COS or subcritical baroclinic instability (SBI; e.g. \citealt{klahr2003, lesur2010}. The viscosity still increases monotonically with $\chi_p$—with the sole exception of the $\chi_p=4\times 10^{-5}$ case. The origin of this deviation is not entirely clear, though it is almost certainly connected to the interaction between the inward-migrating vortices and the planet.

Taken together, these results indicate that the effective turbulent viscosity generated by vortices does not behave like a laminar Navier–Stokes viscosity—indeed, one would not expect it to.

We note that we do not find clear evidence for de-saturation of the corotation torque in either 2D or 3D simulations. In 2D, this is unsurprising because the simulations are inviscid and maintaining the entropy- and vortensity-related corotation torques requires a sufficiently large viscosity. In 3D, the situation is more complex: COS modifies the disk structure and thereby alters the entropy and vortensity gradients in a non-trivial way, making de-saturation difficult to assess.

\subsection{Vortex-planet interaction}

To better understand the encounter between the planet and migrating vortices, let us consider the torque exerted by the planet on the vortex. Let the planet of mass $M_p$ be fixed at a position $\vec{r}_p$ and the vortex of mass $M_v$ be at $\vec{r}_v$. The torque exerted by the planet on the vortex, calculated about the central star, is given by the $z$-component of $\vec{\Gamma}_{p \to v} = \vec{r}_v \times \vec{F}_{p \to v}$, where $\vec{F}_{p \to v}$ is the gravitational force of the planet on the vortex.

If we place the planet on the positive x-axis at $\vec{r}_p = (r_p, 0)$ and assume orbital motion is in the positive $\phi$ direction (counter-clockwise), the torque simplifies to:
\begin{equation}
 \Gamma_{p \to v} = - \left( \frac{G M_p M_v}{|\vec{r}_v - \vec{r}_p|^3} \right) r_p r_v \sin\phi_v    
\end{equation}
The sign of the instantaneous torque depends entirely on the vortex's azimuthal position relative to the planet, $\phi_v$.

Vortices typically form in the outer disk ($r_v > r_p$) where the planet's gravitational influence is initially negligible. These vortices migrate inward due to the excitation of their own spiral density waves \citep{paardekooper2010,lehmann2024}. As a vortex approaches the planet, its angular velocity is lower than that of the planet. Consequently, in the frame of the planet, the vortex approaches from the leading direction ($\phi_v > 0$). From this initial geometry, our simulations reveal three distinct interaction outcomes depending on the vortex's impact parameter and the local flow dynamics.

First, we observe cases where vortices migrate radially across the planet's orbit ($r > r_p$ to $r < r_p$) without generating a significant torque signature. These crossings do not exhibit the kinematics of a gravitational scattering event. We speculate that these are driven by background hydrodynamic torques—such as interactions with spiral waves from other vortices or the COS turbulence—which effectively push the vortex past the co-orbital barrier.

In a second type of passage, the vortex avoids a close scattering encounter and drifts past the planet azimuthally. As it traverses from the leading side ($\phi_v > 0$) to the trailing side ($\phi_v < 0$), the torque it exerts on the planet reverses sign from positive to negative. Our results show instances where this passage results in a net negative torque "kick" (e.g., the simulation with $\chi_p = 2 \cdot 10^{-5}$ in Figure \ref{fig:torque_cumu} (top panel) at $\approx 300$ orbits). This suggests that the vortex may linger in the trailing region, potentially due to a hydrodynamic interaction with the planet's spiral wake, which trails the planet and could act to slow the vortex's relative azimuthal drift.

Finally, the third type of observed passage is the one crucial for the large positive torque events (seen in many of the curves in Figure \ref{fig:torque_cumu})—the vortex undergoes a distinct horseshoe interaction. Instead of passing to the trailing side, the vortex is scattered by the planet's gravity while still in the leading region. It executes a U-turn that transfers it from its initial outer orbit to an inner orbit ($r < r_p$) downstream of the planet. Because this entire interaction occurs while the vortex is located at $\phi_v > 0$, the planet consistently exerts a negative torque on the vortex. By conservation of angular momentum, the torque exerted by the vortex on the planet is strictly positive ($\Gamma_{v \to p} > 0$), explaining the rapid upward "kicks" in the migration tracks.

\section{Discussion}\label{sec:discussion}

\subsection{Thermal Diffusion implementations}

We find a clear positive correlation between the thermal diffusivity $\chi_p$ and the strength of COS-induced turbulence only when using prescription A, which diffuses entropy (Figure \ref{fig:COS_sat_tdiff}). This trend also manifests in the corresponding COS-driven modification of the planet torque and, ultimately, in the inferred migration timescale (Figure \ref{fig:migration}).

In contrast, prescriptions B and C, which diffuse temperature rather than entropy, show no clear correlation between $\chi_p$ and the turbulence amplitude. Because the turbulence is primarily driven by vortex-launched spiral density waves, its strength directly traces the presence and persistence of large vortices.

One possible explanation for the differing behavior among the diffusion prescriptions is that the SBI may most effectively amplify large vortices for intermediate diffusivities of $\chi_p \sim 5\times10^{-6}$. At these values, turbulent viscosities are largest, and indeed we see a significant enhancement of migration timescales between 2D and 3D simulations for prescriptions B and C.


Other effects could in principle influence vortex survival. For example, the elliptic instability \citep{lesur2009}, or “thermal” damping of vortex-induced spiral waves (Section \ref{sec:damping}) might play a role, but the latter should become relevant only at much larger values of $\chi$ than those explored here. Cooling-induced baroclinic spin-down of vortices \citep{fung2021} is also unlikely to operate in our parameter regime. Their analysis shows that this mechanism is most effective when the cooling time is slightly smaller than the vortex turnover time.

The turnover time for an elliptical vortex in a Keplerian disk is \citep{lesur2009}
\begin{equation}
t_{\mathrm{turn}} = \frac{4\pi (\mathcal{A}-1)}{3},
\end{equation}
where $\mathcal{A}$ is the vortex aspect ratio. For typical values $\mathcal{A} \sim 5$--$10$ \citep{lehmann2024}, we obtain $t_{\mathrm{turn}} \sim 20$--$40$ orbital periods.

The effective cooling time can be estimated by taking a vortex width of one scale height $H$. For thermal diffusivities $\chi \sim 5\times10^{-6}$--$4\times10^{-5}$, this yields
$t_{\mathrm{cool}} \sim \frac{H^2}{\chi} \sim 500\text{--}2000$ orbital periods,
which is much longer than $t_{\mathrm{turn}}$. This places our simulations far outside the regime where baroclinic vortex spin-down is expected to be significant, according to \citet{fung2021}.

On the numerical side, we observe a modest loss of thermal energy at the inner boundary of the computational domain in the 3D simulations that develop COS. In runs employing thermal diffusion prescriptions B and C, this leads to a mild flattening of the temperature profile toward smaller radii, including the innermost region. A flatter temperature gradient would, in principle, weaken the COS. In contrast, simulations using prescription A show a slight increase in temperature near the inner boundary, which could locally enhance the COS. However, since the vortices that primarily affect the planetary torque form exterior to the planet and migrate inward past it, we do not expect these boundary-related temperature variations to significantly influence our main results.

\subsection{Representative PPD models}\label{sec:ppdmodel}

To apply our findings to PPDs we first calculate radial profiles of the radiative thermal diffusivity, \(\chi(r)\), for three representative PPD models. These are an observationally-motivated, low-mass PPD model, the classical Minimum Mass Solar Nebula (MMSN, \citealt{hayashi1981}), as well as a massive disk model, and use the same models to estimate physical planet migration time scales from our measured simulation torques in Section \ref{sec:migration} below.

For the low mass disk model we adopt a gas surface density corresponding to $0.1\times$ the MMSN. To place this low-mass disk in a consistent observational context, we assume a central star of mass $M_*=0.5\,M_\odot$. This choice is motivated by large-scale ALMA surveys, which find a steep correlation between disk and stellar mass ($M_{\rm dust} \propto M_*^{1.3-1.9}$, \citealt{Pascucci2016}). Specifically, the median dust mass for a $0.5\,M_\odot$ star is found to be $\sim 5-6\,M_\oplus$; assuming a standard gas-to-dust ratio of 100, this yields a gas mass of $\sim 0.0015\,M_\odot$. Furthermore, this stellar mass naturally accounts for the cooler thermal structure in our model: assuming passive irradiation ($T \propto L_*^{1/4} \propto M_*$)\footnote{Here $L_*$ denotes the stellar luminosity. We assume a standard main-sequence mass-luminosity relationship $L_* \propto M_*^4$, such that $T \propto (M_*^4)^{1/4} \propto M_*$.}, a $0.5\,M_\odot$ star results in disk temperatures roughly half that of a solar-analog MMSN (e.g., $\sim 150$\,K vs $280$\,K at 1\,AU).

Based on this, we define our low mass disk model with the following profiles for surface density \(\Sigma\) and midplane temperature \(T\):
\begin{align}
    \Sigma(r)& =170\,(r/\mathrm{AU})^{-0.5}\,\mathrm{g\,cm^{-2}}\label{eq:model_sigma},\\
     T(r)& = 150\,(r/\mathrm{AU})^{-1.0}\,\mathrm{K}\label{eq:model_temp}.
    \end{align}
The surface density normalization at 1 AU is 10\% of the MMSN \citep{hayashi1981}, and the temperature normalization is significantly reduced from the canonical 280 K. The power-law exponents, a surface density slope of \(p=0.5\) and a temperature slope of \(q=1.0\), are chosen to match the setup of our hydrodynamic simulations.


\begin{figure*}
    \centering
    \includegraphics[width=1.0\textwidth]{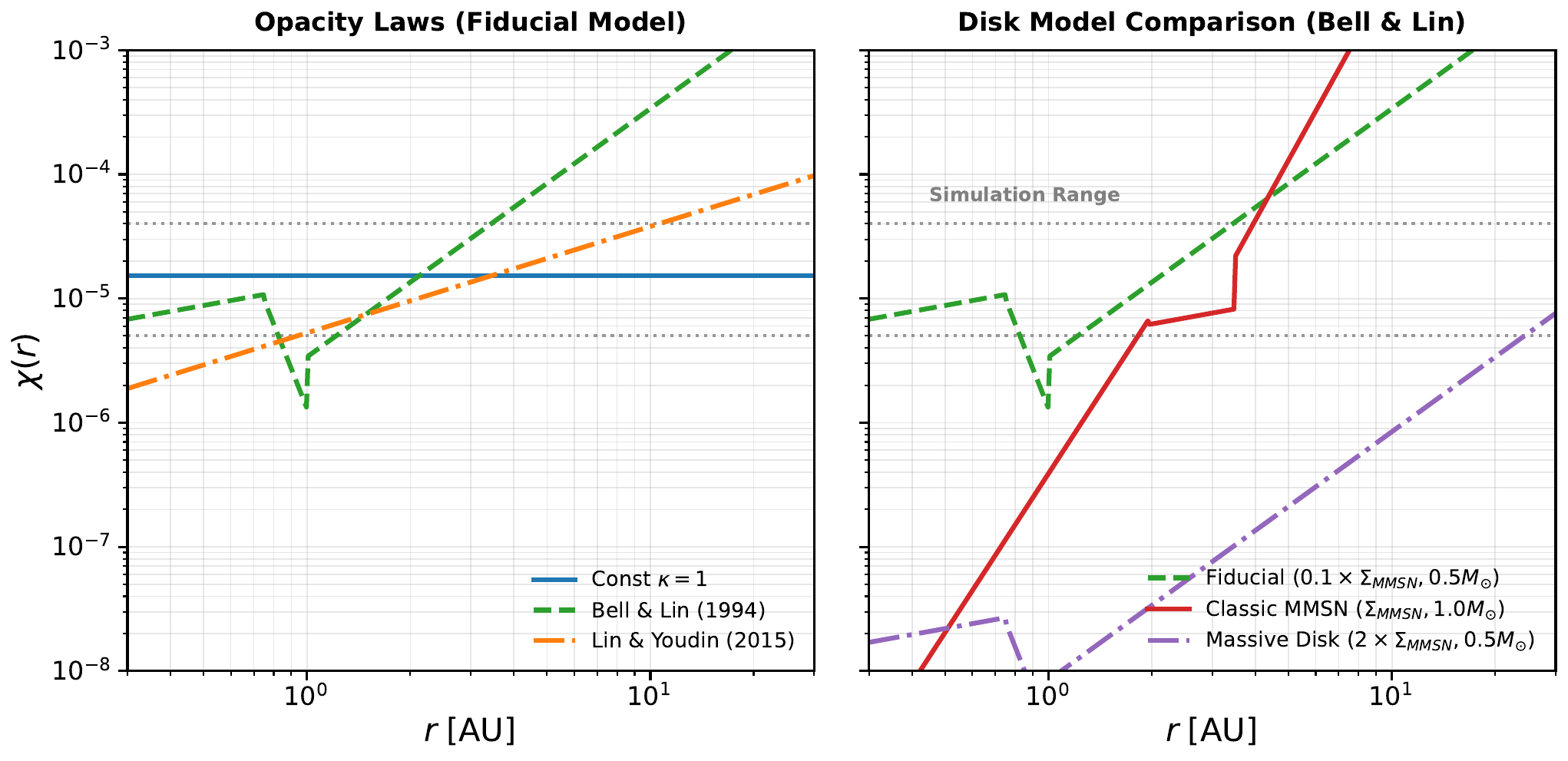} 
    \caption{ \textbf{Left:} Radial profiles of the dimensionless midplane thermal diffusivity $\chi$ (normalized by $r_0^2 \Omega_K(r_0)$ at $r_0=1$ AU) for our fiducial low-mass disk model ($0.1 \times \Sigma_{\text{MMSN}}$, $M_*=0.5 M_{\odot}$) using three different opacity prescriptions: constant opacity (blue), Bell \& Lin (1994) (green dashed), and Lin \& Youdin (2015) (orange dash-dotted). 
    \textbf{Right:} Comparison of $\chi$ profiles using the Bell \& Lin (1994) opacity for three different disk models: our fiducial model ($p=0.5, q=1.0$; green dashed), a standard MMSN disk around a $1.0 M_{\odot}$ star ($p=1.5, q=0.5$; red solid), and a massive disk ($2 \times \Sigma_{\text{MMSN}}$) around a $0.5 M_{\odot}$ star ($p=0.5, q=1.0$; purple dash-dotted). 
    The horizontal dotted lines mark the range of thermal diffusivities used in our simulations ($\chi \in [5\times10^{-6}, 4\times10^{-5}]$). The local peak in the Bell \& Lin diffusivity profiles at $r \approx 1$--$3$ AU corresponds to the water ice sublimation front (the ice line), where the opacity drops abruptly due to the evaporation of ice grains, leading to a localized increase in thermal diffusion.}
    \label{fig:chi_profile}
\end{figure*}

The physical thermal diffusivity in the optically thick midplane of a PPD is given by \citep{mihalas1984}
\begin{equation}
    \chi \;=\; \frac{K}{\rho c_{P}},
\end{equation}
where \(\rho\) is the midplane gas density, \(c_{P}\) is the specific heat at constant pressure, and \(K\) is the radiative conductivity. For a gas of ideal molecules with mean molecular weight \(\mu\) and adiabatic index \(\gamma\), we have \(c_{P} = \frac{\gamma}{\gamma-1}\,\frac{k_{\mathrm{B}}}{\mu m_{\mathrm{H}}}\). The conductivity is \(K = \frac{16\,\sigma_{\mathrm{SB}}\,T^{3}}{3\,\kappa\,\rho}\), where \(T\) is the midplane temperature, \(\kappa\) is the Rosseland mean opacity, $k_{\text{B}}$ the Boltzmann constant, and \(\sigma_{\mathrm{SB}}\) is the Stefan-Boltzmann constant. Combining these and using the midplane density approximation \(\rho=\Sigma/(\sqrt{2\pi}\,H)\) where \(H=c_s/\Omega\) is the disk scale height, we can express the diffusivity as:
\begin{equation}
    \chi(r) = \frac{32 \pi \sigma_{\mathrm{SB}} (\gamma-1)}{3 G M_{*} \gamma} \frac{T(r)^4 r^3}{\kappa(r) \Sigma(r)^2}.
\end{equation}
Recall the dimensionless diffusivity used in our simulations is defined with a fixed normalization at $r_p$ with $\chi \to \chi / (r_p^2 \Omega_p)$, where $\Omega_p = \Omega(r_p)$.

For our calculations, we assume an ideal gas with \(\mu=2.34\) and \(\gamma=1.4\). We evaluate \(\chi(r)\) for three opacity prescriptions:
\begin{enumerate}
    \item A constant opacity \(\kappa=1\,\mathrm{cm^{2}\,g^{-1}}\)\citep{benitez2015}.
    \item A piecewise, dust-dominated opacity law from \citet{bell1994}, implemented as
    \[
        \kappa(T)=
        \begin{cases}
            2\times10^{-4}\,T^{2} & (T<150\,\mathrm{K}),\\
            2\times10^{16}\,T^{-7} & (150\le T<200\,\mathrm{K}),\\
            0.1\,T^{1/2} & (T\ge 200\,\mathrm{K}),
        \end{cases}
    \]
    with \(\kappa\) in \(\mathrm{cm^{2}\,g^{-1}}\) and \(T\) in K. 
    \item The dust opacity from \citet{lin2015} (their Eq.~58),
    \begin{equation}
        \kappa_{d}(r) \;=\; 2.88\,\hat{\kappa}_{d}\,
        \left(\frac{r}{\mathrm{AU}}\right)^{-6/7}\,\mathrm{cm^{2}\,g^{-1}},
    \end{equation}
    where we adopt the fiducial dust abundance scaling \(\hat{\kappa}_{d}\!=\!1\).
\end{enumerate}
 The piecewise opacity function captures the water ice sublimation transition at temperatures between $150-200$ K. 
 
For our fiducial disk with $p=-0.5$ and $q=-1.0$, the radial dependencies in the formula for the physical diffusivity conveniently cancel, leaving it primarily dependent on the local opacity: $\chi(r) \propto 1/\kappa(r)$. Figure \ref{fig:chi_profile} shows the resulting dimensionless thermal diffusivity profiles. The left panel illustrates the sensitivity to different opacity prescriptions for our fiducial model. The right panel contextualizes this choice by comparing the fiducial disk against a standard Minimum Mass Solar Nebula (MMSN, $M_*=1.0 M_{\odot}$) and a more massive disk model ($2\times\Sigma_{\text{MMSN}}$, $M_*=0.5 M_{\odot}$). We find that the range of thermal diffusivities used in our simulations ($\chi \in [5\times10^{-6}, 4\times10^{-5}]$) is physically robust across these environments. In both the fiducial low-mass model and the standard MMSN, these diffusivities correspond to the region outside of the water ice line region ($r \sim 1-4$ AU), where planet formation is thought to be efficient (e.g., \citealt{bitsch2015, danti2025}). In the massive disk model, these values map to larger radii ($r \sim 20-30$ AU).

\subsection{Prospective migration time scales}\label{sec:migration}

\begin{figure}
    \centering
    \includegraphics[width=0.5\textwidth]{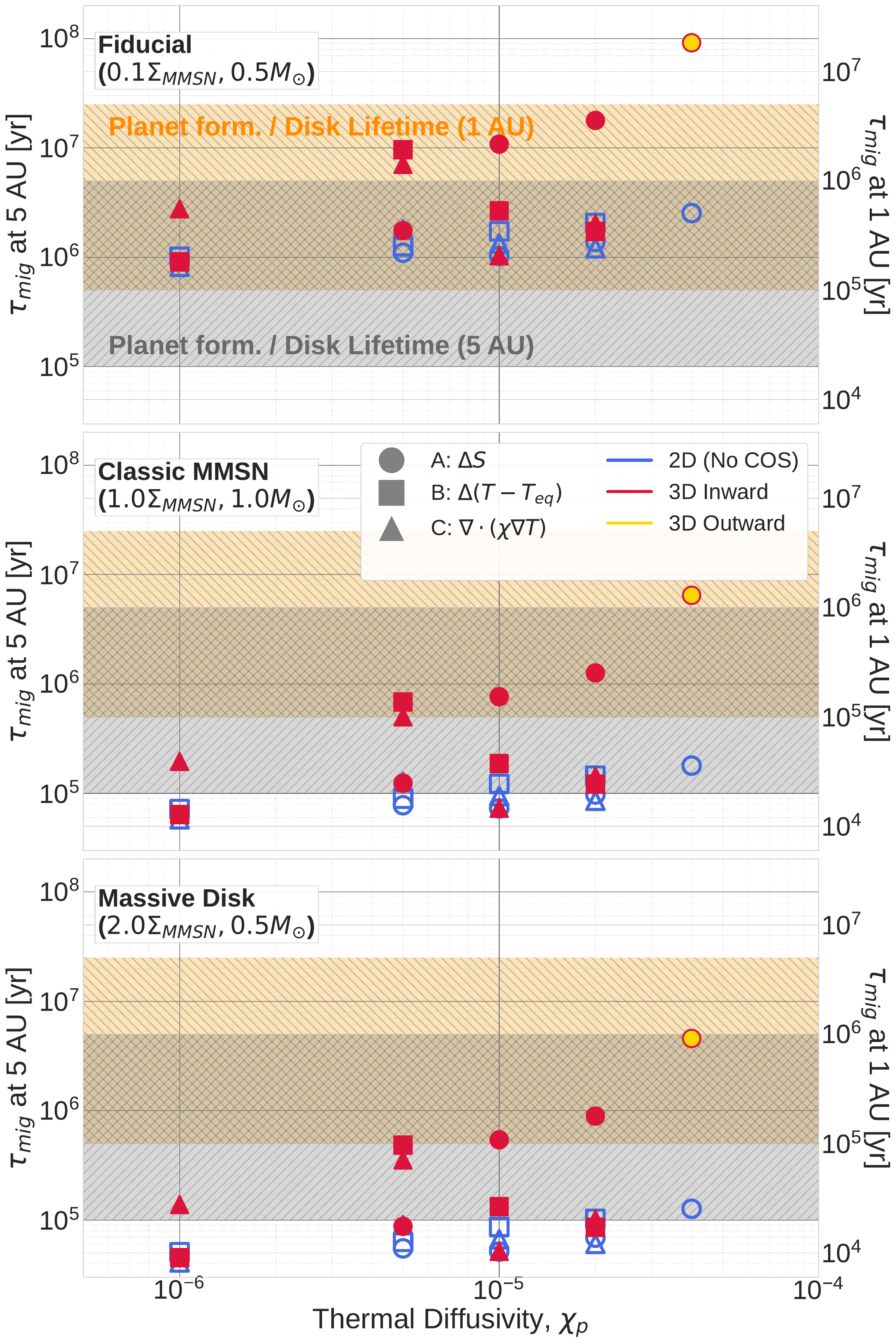} 
    \caption{Prospective migration timescales of a nascent low-mass planet ($q=2.52 \cdot 10^{-5}$) plotted against thermal diffusivity $\chi$. The three panels correspond to different disk models: \textbf{Top:} Fiducial low-mass disk ($0.1 \Sigma_{\text{MMSN}}$); \textbf{Middle:} Classic MMSN ($1.0 \Sigma_{\text{MMSN}}$); \textbf{Bottom:} Massive disk ($2.0 \Sigma_{\text{MMSN}}$). In all panels, the left y-axis shows the timescale at 5 AU, while the right y-axis shows the corresponding timescale at 1 AU. Solid red markers denote 3D simulations with COS, while hollow blue markers denote 2D reference runs. Yellow markers indicate cases of outward migration. The hatched bands mark the typical range of super earth formation time and protoplanetary disk lifetimes ($0.1$--$5$ Myr).}
    \label{fig:migration}
\end{figure}

The raw torques from our simulations can be used to compute a rough estimate for the migration timescale:
\begin{equation}\label{eq:migration}
    \tau_a \equiv a/\langle \dot{a}\rangle_{t},
\end{equation}
where $a$ is the planet's current location and $\dot{a}$ is the migration rate resulting from the time-averaged torque. Using Eq. (\ref{eq:ode}) we find
\begin{equation}
    \tau_a = \frac{\sqrt{G M_{*}a}}{2 \langle{\hat{\Gamma}\rangle_{t}}}.
\end{equation}

Since our simulations are scale-free, we can interpret them in the context of arbitrary locations within a protoplanetary disk, provided the applied physics are valid there. In particular, since COS is expected to occur at radii of $\sim 1-10$ AU \citep{pfeil2019}, we consider here the two radii $r_p=1$ AU and $r_p=5$ AU. The time scale $\tau_a$ of planet migration depends on the physical value of the disk's surface mass density via the torque. All simulations use a fixed dimensionless surface density of $\Sigma_p = 6.366 \times 10^{-4}$ in code units (Section \ref{sec:numerics}) at their local reference radius $r_p$. For a central star of mass $M_*=0.5 M_\odot$, this corresponds to a physical surface density of $\Sigma_{\mathrm{sim}} \approx 2828 \, \mathrm{g \, cm^{-2}}$ at 1 AU, and $\Sigma_{\mathrm{sim}} \approx 113 \, \mathrm{g \, cm^{-2}}$ for the 5 AU interpretation.

To provide a physically meaningful comparison with planet formation theory, we scale the resulting timescales. First, to ensure a fair comparison between our 2D and 3D simulations, the migration timescales from the 3D simulations are internally recalibrated to the physical scale of their 2D counterparts at each location. Second, both the 2D and re-calibrated 3D timescales are scaled to the PPD disk models presented above. For instance, the low mass disk model has a surface density given by Eq. (\ref{eq:model_sigma}). The final timescales presented in our results are scaled using the relation $\tau_{\mathrm{final}} = \tau_{\mathrm{sim}} \times (\Sigma_{\mathrm{sim}} / \Sigma_{\mathrm{model}})$, evaluated independently at 1 AU and 5 AU.

Figure \ref{fig:migration} presents the estimated migration timescales $|\tau_a|$ as a function of thermal diffusivity $\chi$ across the same three disk models used in Figure \ref{fig:chi_profile}: our fiducial low-mass disk (top panel), the standard MMSN (middle panel), and the massive disk (bottom panel). In each panel, the left vertical axis indicates the timescale for a planet at 5 AU, while the right vertical axis shows the corresponding timescale for a planet at 1 AU. The latter is notably shorter due to the combined effects of the higher local gas surface density and the shorter orbital period at 1 AU. Solid red symbols show results from 3D simulations where COS is active, while hollow blue symbols show results from corresponding 2D simulations without COS. The marker shapes (circles, squares, triangles) correspond to the different thermal diffusion methods described in Section \ref{sec:thermal}. The gray and orange hatched regions indicate the canonical timescales for giant planet core formation and typical disk lifetimes ($0.1-5$ Myr; e.g., \citealt{mamajek2009, bitsch2019, danti2025}).

At both distances, COS extends migration timescales by up to approximately a factor of 10 compared to the laminar 2D case. Comparing the three panels, the migration timescales are roughly an order of magnitude shorter in the Classic MMSN and Massive disk models compared to the fiducial case. This acceleration is primarily driven by the higher gas surface density ($\Gamma \propto \Sigma$), though the different stellar masses also contribute by altering the orbital period ($P_{orb} \propto M_*^{-1/2}$).
We find that for our fiducial low-mass model (top), the migration timescales generally fall within or above the disk lifetime, suggesting survival is likely. However, for the denser MMSN and massive disk models (middle and bottom), the timescales drop significantly (in many cases $|\tau_a| < 10^5$ yr), indicating rapid inward migration that poses a challenge for planet survival. Note that most simulations yield inward migration, with one exception: the simulation with thermal diffusion method A and $\chi_p = 4 \cdot 10^{-5}$ yields a \emph{positive} time-averaged torque and hence \emph{outward} migration. This case is indicated by a solid yellow circle.

\subsection{Comparison with previous works}

\subsubsection{Planet-Disk interactions}

The torques in our 3D simulations are highly oscillatory, undergoing stochastic changes, in agreement with the simulations of \citet{stollpicogna2017}, that developed the VSI. 
However, in that study, vortices formed only via RWI at edges of gaps carved by massive planets. No vortices occurred in their low-mass planet simulations.

Our finding that COS-induced vortices that traverse the planet in most cases provide a net positive torque---an "outward kick"--- may at first seem to contradict the results of previous studies, such as \citet{faure2016}, who found that vortices tend to drag planets inward. However, here we explore a fundamentally different physical regime of planet-vortex interaction. The discrepancy in outcomes stems from key differences in the origin and, most critically, the mass of the vortices relative to the planet.

In the scenario modeled by \citet{faure2016}, vortices are generated by the RWI at the sharp density maximum of a dead zone inner edge. This process creates vortices that are an order of magnitude more massive than the embedded planet. In this high-mass-vortex regime, the interaction is one of gravitational capture; the planet's dynamics are dominated by the vortex, which captures the planet in its potential well and drags it along its own inward migration path.

In contrast, our work investigates the self-consistent generation of vortices by the COS. A key result of our simulations is that these vortices are significantly less massive than the planet itself; for instance, in Appendix \ref{app:vortex_mass} we calculate the mass of a vortex providing a significant positive torque kick to be just $0.36 M_p$. In this low-mass-vortex regime, the planet is the gravitationally dominant body. The interaction is not capture, but rather a scattering event. The less-massive vortex can be forced onto a horseshoe orbit relative to the more-massive planet, and in the process of executing its U-turn from an outer to an inner orbit, it transfers angular momentum to the planet. This transfer manifests as the positive torque we measure.

By utilizing a fixed planetary orbit, our study isolates this fundamental torque exchange, a necessary first step before tackling the more complex dynamics of a fully migrating planet. Our findings therefore extend the understanding of planet-vortex interactions, demonstrating that the outcome---inward drag versus an outward kick---is highly sensitive to the mass ratio of the system. Our work highlights a new and physically motivated mechanism by which the turbulence inherent in a COS-active region can counteract inward planet migration.

\subsubsection{Saturation of the COS}
Our adopted values of thermal diffusivity $\chi_p \sim 10^{-6}-10^{-5}$ with an aspect ratio $h_p=0.1$ correspond to Peclet numbers $Pe\equiv \frac{H_p^2 \Omega_p}{\chi_p}\sim 10^3- 10^4$. Based on the results of \citet{teed2025}, we would expect only weak non-axisymmetries, and therefore a small turbulent viscosity, which contrasts our findings. However, note that our simulation are inviscid, implying a very large Reynolds number, much larger than probed by \citet{teed2025}.
We ran one additional 3D simulation with thermal diffusion model B and a viscosity $\nu_p=10^{-7}$, which corresponds to a Reynolds number $Re \equiv \frac{H_p^2 \Omega_p}{\nu_p}=10^5$ as used by \citet{teed2025}, and this simulation showed no formation of vortices, and correspondingly, very small turbulent viscosity values within 1000 orbits.

The lack of vortices in our optically thin cooling simulations contrasts with the simulations of \citet{lehmann2024,lehmann2025}, who found the development of large vortices under optically thin cooling. 
One possible explanation for the absence of vortices in the $\beta$ cooling simulations could be that the Rossby wave instability, which ultimately breaks up zonal flows into vortices is expected to be weaker for beta-cooling than for thermal diffusion for the current parameter regime. This is evident from Figures 8 and 9 of \citet{huang2022}. 
The finding of large vortices in the simulations of \citet{lehmann2024,lehmann2025}  likely stems from the steeper temperature slope ($q = 2$) used in those studies, which significantly amplifies COS activity compared to our more moderate slope ($q = 1$) used here.

\subsubsection{Migration Time scales}

It is instructive to contrast our predicted migration time scales with recent works on pebble-accretion planet formation models, such as \citet{danti2025}, which find limited inward migration for forming super-Earths, suggesting that no additional mechanisms are required to safe planets from being lost to the star. Their outcome, however, is largely a consequence of their adopted disk model, which features a large effective viscosity ($\alpha \sim 10^{-2}$). This high viscosity drastically alters the migration landscape in two ways. First, it generates significant viscous heating, creating a hot, low-density inner disk that naturally suppresses Type I migration rates (which scale as $\Sigma (H/r)^{-2}$). Second, and perhaps more critically, this high viscosity implies a significant desaturation of the corotation torque. By adopting a fixed migration coefficient ($c=2.8$) based on \citet{paardekooper2010}, \citet{danti2025} implicitly assume a fully efficient, unsaturated corotation torque. In a low viscosity environment characteristic of dead zones, this torque would (at least partially) saturate, and potentially be outweighed by the inward Lindblad torque. Furthermore, their models invoke a transition to slower Type II-like migration once planets approach pebble-isolation mass. It is unclear if such gap-induced slowing is physically consistent with their high-$\alpha$ regime, where viscous refilling typically prevents gap opening for low-mass planets. In contrast, our simulations represent inviscid, magnetically quiet regions. Moreover, we do not find significant signs of gap opening in our simulations, indicating that gap-induced migration slow-down is negligible in our setup. Instead, our work isolates a complementary hydrodynamic mechanism—COS-driven turbulence—that can prolong the survival of few-Earth-mass planets in environments where rapid type-I migration would otherwise be expected.

\subsection{Caveats}\label{sec:caveats}



The assumed fiducial low mass disk model in Section \ref{sec:ppdmodel}, based on passive stellar irradiation, implies a midplane aspect ratio of $h(r)=H/r \approx 0.025$. The aspect ratio used in our hydrodynamic simulations ($h_p=0.1$) is intentionally set to a larger value. This choice is motivated by our goal to study the saturated state of the COS in a vertically extended disk, which is both physically plausible and numerically tractable. 
Therefore, our choice of $h_p=0.1$ should be viewed primarily as a feature of our numerical experiment that allows us to study the relevant physics within reasonable computational limits. A larger scale height accelerates the non-linear saturation of the instability, reducing the required simulation time. Furthermore, a smaller aspect ratio, e.g. $h_p \approx 0.025$, would require prohibitively high grid resolution. Note also that, strictly speaking, the classic MMSN disk model is actually stable to the COS based on the slopes $p=1.5$ and $q=0.5$.

Our simulations measure the instantaneous torque on a planet held at a fixed orbital radius for 1,000 orbits (approximately 1.4 kyr at 1 AU). 
This allows for direct comparison to the comprehensive model of \PK, which strictly applies to non-migrating planets. If we allowed the planet to migrate freely, the results could be different. For example, dynamical co-rotational torques then become possible \citep{paardekooper2014}. 
However, the main mechanism of angular momentum transfer between the planet and inward migrating vortices of significantly lower mass should still hold, provided the planet's migration rate does not exceed the vortices' migration rates. 
Note also that the inferred migration time scales ($10^5$--$10^8$ years) far exceed the simulation duration ($t_{\rm sim} \ll \tau_a$), validating the fixed-orbit approach. Furthermore, when $\tau_a$ exceeds typical super-Earth formation timescales, the migration is slow enough that the planet can be considered dynamically stable throughout its formation process. Thus, the COS regions in these regimes act as effective ``safe havens,'' delaying inward drift long enough for the planet to fully assemble.

We also keep the planet mass fixed, i.e. we assume the planet has its final pebble isolation mass throughout the entire simulation. In principle, the planet would start with its planetesimal mass and grow via pebble accretion \citep{lambrechts2014, danti2025}.
Our assumption of a fixed planetary mass serves as a stringent test of the above described survival criterion. Since Type I migration rates generally scale with planet mass in the linear low-mass regime ($\dot{a} \propto M_p$), the fully grown planet experiences the fastest migration rate of its formation history. Furthermore, in the pebble accretion paradigm, the growth timescale is typically significantly shorter than the global disk lifetime \citep{lambrechts2014, bitsch2015}, meaning the planet spends the majority of its embedded phase at its final (pebble isolation) mass. By demonstrating that COS-driven turbulence can extend $\tau_a$ beyond the regime of formation timescales for this final mass, we effectively place an upper bound on the inward migration efficiency; if the planet survives this ``fastest-drift'' phase, it is unlikely to have been lost during the briefer, lower-mass accretion stages.

While our simulations are 3D, they ignore vertical disk stratification. If COS-induced turbulence behaves similarly throughout the entire vertical domain, then these results are expected to hold qualitatively. However, if COS vortices were concentrated to the midplane, then the stable gas away from the midplane could overwhelm this effect and push the results back into the laminar regime.
However, we confirmed that the COS vortices here are vertically unstructured, with their columns extending throughout the entire vertical domain (cf. Figure \ref{fig:COS_sat_tdiff_2}), which lends credibility to our results. 

In principle, additional complications are expected to result from vertical stratification. For instance, the VSI may occur and interplay with the COS. Exploring the nonlinear interaction between vertically global VSI-driven turbulence and COS vortices remains an important open problem which should be addressed in future work.

Furthermore, vertical stratification gives rise to vertical buoyancy.
The latter is not expected to directly weaken the COS in a significant manner \citep{lehmann2023}.
However, it allows the planet to excite internal gravity waves (buoyancy waves) at buoyancy resonances \citep{Zhu2012, McNally2020}. In laminar disks with vertically isothermal profiles typical of stellar irradiation, \citet{Zhu2012} found that these waves exert a direct negative torque roughly $10-40\%$ of the Lindblad torque. More dramatically, \citet{McNally2020} showed that these waves can alter the vortensity of the horseshoe region, potentially reversing the sign of the dynamical corotation torque and leading to rapid inward migration. However, buoyancy waves are subject to damping by thermal diffusion \citep{Zhu2012, yun2022}. Given the finite thermal diffusivity required for the COS in our models, such damping is expected to reduce the efficacy of the buoyancy torque compared to the adiabatic limit. Moreover, the buoyancy-induced reversal of the corotation torque relies on the coherent accumulation of vortensity modifications. As noted by \citet{McNally2020}, this mechanism is likely disrupted by vortex mixing. Consequently, while our unstratified models exclude these negative torque components, they are expected to be sub-dominant compared to the large-amplitude, stochastic torque fluctuations driven by vortex scattering in the COS-active regime.

Finally, we note that the thermodynamics governing the COS also affect the thermal torque \citep{Lega2014, Masset2017}. In the regime of high thermal diffusivity, a non-luminous planet experiences a `cold thermal torque' arising from the diffusion of heat out of compressed gas lobes within the Hill sphere. This effect typically adds a negative torque contribution. We believe, however, that our simulations likely do not resolve this effect, as our potential smoothing length ($b=0.4H$) is larger than the planetary Hill radius.

\section{Summary}

This study investigates how COS-driven turbulence modifies planet-disk torques through comprehensive hydrodynamic simulations.  We employ both thermal diffusion and optically thin cooling prescriptions and validate our methods against the established theoretical framework of \PK. Our approach focuses on fixed, non-migrating planets to isolate the fundamental torque modifications induced by COS turbulence. Crucially, COS is a 3D instability that is absent in 2D simulations, making the comparison between 2D and 3D results particularly revealing. 

Our key findings demonstrate that COS activity fundamentally alters planet-disk interactions, but only under specific thermal conditions. In simulations with thermal diffusion, the instability saturates into large-scale, coherent vortices that migrate radially through the disk. When these vortices encounter an embedded planet, they typically undergo gravitational scattering events that transfer angular momentum to the planet, manifesting as positive torque "kicks" that counteract inward migration.

The physical mechanism underlying this effect stems from the mass ratio between planets and COS-generated vortices. Our simulations show that these vortices are significantly less massive than typical Super-Earth mass planets (vortex masses $\sim 0.1 M_p$), placing the interaction in a scattering regime rather than a capture regime. The planet gravitationally dominates the encounter, forcing the vortex onto a horseshoe orbit and gaining angular momentum in the process.

However, in contrast to thermal-diffusion simulations, our optically thin cooling simulations do not produce significant changes in the torque. Under these conditions, COS saturates in near-axisymmetric structures without forming the large-scale vortices necessary for substantial planet-vortex interactions. This contrasts with the simulations of \citet{lehmann2024,lehmann2025}, who found the development of large vortices under optically thin cooling. The difference likely stems from the steeper temperature slope ($q = 2$) used in those studies, which significantly amplifies COS activity compared to our more moderate slope ($q = 1$). This highlights the importance of realistic thermal modeling and disk structure in understanding planet-disk dynamics. Note that we cannot rule out the possibility that vortex formation could still occur with optically thin cooling at times later than 1000 orbits.

We quantify these effects by computing prospective migration timescales for planets in representative PPD models. Our results indicate that COS activity via thermal diffusion can extend migration timescales by approximately 10, sufficient to overlap with or even exceed typical Super-Earth formation and disk lifetime windows (0.1 to 5 Myr). This suggests a viable mechanism for planet survival in the face of rapid Type I migration.

These findings may have important implications for planet formation theory and the interpretation of exoplanet observations. The presence of COS in realistic disk models with appropriate thermal physics, combined with its ability to generate positive torques from vortex-scattering, may provide a natural explanation for the survival of close-in planetary populations without requiring fine-tuned initial conditions or special features such as planet traps.  
Furthermore, our work establishes a framework for understanding planet-turbulence interactions in the context of disk evolution models that emphasize hydrodynamic rather than magnetic processes.

\begin{acknowledgments}
We thank an anonymous referee for providing a clear and constructive report which helped to improve the manuscript.
This work is supported by the National Science and Technology
Council (grants 114-2811-M-001-022-, 114-2112-M-001-018-, 
114-2124-M-002-003-, 115-2124-M-002-014-), an Academia 
Sinica Career Development Award (AS-CDA-110-M06), and an Academia Sinica Grand Challenge Seed Grant (AS-GCS-115-M02).  
Simulations were performed on the \emph{Kawas} cluster at ASIAA and on computing resources provided by Academia Sinica Grid-computing Centre (ASGC).  
We thank ASGC for providing computational and storage resources.
ML also acknowledges support from the National Science Foundation under Grant No.\ 2407762.
\end{acknowledgments}

\appendix

\section{Lindblad torque in presence of optically thin cooling and thermal diffusion}\label{app:gamma_eff_lin}

\subsection{Effective adiabatic index}

The dispersion relation  subject to optically thin $\beta$-cooling can be obtained, e.g. from Eq (29) of \citet{lehmann2024} in the limit $k_z\to 0$ (2D), $\omega\gg \Omega$ (focusing on acoustic waves), applying to perturbations of the form $\exp \left( i \omega r - i k t\right)$. We find, with real frequency $\omega$ and complex wavenumber $k=k_r + i k_i$:
\begin{equation}\label{eq:dispersion}
    \omega^2/k^2 = c_{s}^2 \frac{1+ i \gamma \omega t_c}{1+ i \omega t_c}.
\end{equation}
In analogy to \PK~, we define the dimensionless measure of cooling on the time scale of the considered wave perturbations
\begin{equation}
    Q_{\beta} = \omega t_c.
\end{equation}
To obtain the phase velocity of acoustic waves we consider
\begin{equation}
   v_p = \omega / k_r,
\end{equation}
and define 
\begin{equation}
    \gamma_{\mathrm{eff}} = \frac{v_p^2}{c_s^2},
\end{equation}
where $k_r=Re[k]$.
After some algebra we find
\begin{equation}\label{eq:gammaeff}
    \gamma_{\mathrm{eff}} = \frac{2}{\left( \frac{1 + \gamma Q_{\beta}^2}{1 + \gamma^2 Q_{\beta}^2} + \sqrt{ \frac{1 + Q_{\beta}^2}{1 + \gamma^2 Q_{\beta}^2} } \right)}
\end{equation}
which smoothly connects the phase velocity of sound waves in the adiabatic and isothermal regimes, i.e. we find $v_p\to c_s$ and $v_p\to \gamma c_s$ for $\beta \to 0$ and $\beta \to \infty$, respectively.

To evaluate $\gamma_\mathrm{eff}$, we need to define an effective value of $Q_\beta$, which amounts to an effective wave frequency. Whenever the waves are resonantly excited, we generally have (in the linear regime) $\omega= m \Omega_p$. Following the argumentation of \PK, most of the torque contribution arises from azimuthal mode numbers $m\sim 2/3 h_p$. Thus, we fix
\begin{equation}\label{eq:Q}
    Q_{\beta} = \frac{2 \beta}{3 h_p}.
\end{equation}

\subsection{Linear wave damping}\label{sec:damping}

In addition to modifying the phase velocity, thermal relaxation introduces a phase shift between pressure and density perturbations, rendering the wavenumber $k$ complex. We write $k = k_r + i k_i$, where $k_i$ represents the linear spatial damping rate of the wave amplitude. 

By inverting the dispersion relation (\ref{eq:dispersion}) and rationalizing the denominator, we obtain
\begin{equation}
    k^2 = \frac{\omega^2}{c_s^2} \frac{1 + i Q_\beta}{1 + i \gamma Q_\beta} = \frac{\omega^2}{c_s^2} \frac{(1 + \gamma Q_\beta^2) - i(\gamma - 1)Q_\beta}{1 + \gamma^2 Q_\beta^2}.
\end{equation}
Assuming weak damping ($|k_i| \ll |k_r|$), we approximate $k^2 \approx k_r^2 + 2i k_r k_i$. Identifying the imaginary components allows us to solve for the damping rate:
\begin{equation}
    k_i \approx -\frac{1}{2 k_r} \frac{\omega^2}{c_s^2} \frac{(\gamma - 1)Q_\beta}{1 + \gamma^2 Q_\beta^2}.
\end{equation}
Substituting the real part $k_r^2 \approx (\omega^2/c_s^2) (1+\gamma Q_\beta^2)/(1+\gamma^2 Q_\beta^2)$, we arrive at
\begin{equation}\label{eq:damping_rate}
    k_i \approx -\frac{k_r}{2} \frac{(\gamma - 1)Q_\beta}{1 + \gamma^2 Q_\beta^2}.
\end{equation}
This expression is consistent with the damping rate of planet-induced spiral density waves derived by \citet{miranda2020}. It indicates that damping vanishes in both the isothermal ($Q_\beta \to 0$) and adiabatic ($Q_\beta \to \infty$) limits, reaching a maximum at intermediate cooling times where $Q_\beta \sim \gamma^{-1/2}$. Strongest damping occurs for 
\begin{equation}
    \beta_{\text{max. damp.}} = \frac{3 h_p}{2 \sqrt{\gamma}}
\end{equation}
yielding 0.12 for $\gamma=1.4$ and $h_p=0.1$
This linear damping mechanism is responsible for the decay of angular momentum flux as waves propagate away from the planet in disks with intermediate cooling timescales and results from out-of-phase oscillations of pressure and density. 

In the case of thermal diffusion, one can similarly derive (assuming weak damping $|k_i|\ll |k_r|$) from Eq (43) of \PK the damping rate
\begin{equation}
    k_i \approx -\frac{k_r}{2} \frac{\left(\gamma-1\right) Q_{\chi}}{1 + Q_{\chi}^2}.
\end{equation}
where 
\begin{equation}\label{eq:qchi}
Q_{\chi} = \frac{2 \chi_p}{3 h_p^3 r_p^2 \Omega_p}.    
\end{equation}
Maximum damping occurs for a thermal diffusion where $Q_{\chi}=1$, such that
\begin{equation}
    \chi_{p,\text{max. damp.}} \approx \frac{3}{2} h_p^3 r_p^2 \Omega_p = 1.5 \cdot 10^{-3} 
\end{equation}
in code units if $h_p=0.1$.

\section{Corotation Torque in presence of optically thin cooling, thermal, and viscous diffusion}\label{app:corot}

\subsection{Saturation}
Co-rotation torques are subject to saturation. In adiabatic inviscid disks, entropy is materially conserved. As co-orbital fluid elements traverse horseshoe orbits, crossing from inner to outer orbits and vice versa, the co-orbital region undergoes phase mixing, entropy gradients vanish, and the related co-orbital torques are erased. Likewise, in barotropic inviscid disks, vortensity (vorticity divided by surface density) is materially conserved, causing the vortensity-related co-rotation torque to saturate. Therefore, dissipative effects, such as viscosity and cooling, are needed to maintain nonlinear co-rotation torques.

To effectively model the corotation torque in the presence of $\beta$-cooling, we must understand the relevant physical timescales, particularly the libration time, and how thermal effects and viscosity influence torque saturation and its transition to the linear regime.

\subsection{Horseshoe half-width}\label{xs_discussion}

The horseshoe region refers to the area around the planet's corotation radius where fluid elements librate instead of circulating past the planet, represented by the blue and red trajectories in Figure \ref{fig:streamlines}. The horseshoe half-width 
  represents the radial distance from the planet where the kinetic energy of the unperturbed Keplerian shear flow is balanced by the effective potential energy barrier in the corotating frame. This barrier is a combination of the planet's direct gravitational potential, depending on the planet mass ratio $q_p$, gravitational softening $b$, disk aspect ratio $h$, as well as the responding gas pressure gradient, the strength of which is determined by the gas thermodynamics encapsulated in the effective adiabatic index, $\gamma_{\text{eff}}$. 

 For low-mass planets, the dimensionless horseshoe half-width $x_s$ is approximately:
\begin{equation}
x_s \approx \frac{1.1}{\gamma_{\text{eff}}^{1/4}} \left(\frac{0.4}{b/h_p}\right)^{1/4}\sqrt{q_p/h_p}
\end{equation}
\citep{paardekooper2010}, which measures the radial size of the region dominated by non-linear horseshoe dynamics relative to $r_p$. 

For our parameters ($q_p=1.26 \times 10^{-5}, h_p=0.05, b/h_p=0.4$) and using $\gamma_{\text{eff}} \approx \gamma = 1.4$, we find $x_s \approx 0.016 < h_p$. It should be noted that $x_s$ depends on gas cooling and the disc aspect ratio via $\gamma_{\text{eff}}$, which depends on $Q_\beta$ as given by Eq (\ref{eq:Q}).

\subsection{Libration time}

The horseshoe libration time $t_{\text{lib}}$ is estimated by considering the fluid motion relative to the planet orbiting at angular velocity $\Omega_p$ and radius $r_p$. The relative azimuthal velocity between gas at radius $r = r_p(1+x)$ and the planet is driven by Keplerian shear, $|\Delta v_\phi| \approx r_p |\Omega_K(r) - \Omega_p| \approx \frac{3}{2} \Omega_p |x| r_p$. Evaluating this speed at the characteristic horseshoe half-width $x_s$, we can find the time to traverse the horseshoe path. One leg of this path covers the full azimuthal distance of $2\pi r_p$ in the co-rotating frame. The time required for this single leg is $t_{\text{leg}} \approx \frac{2\pi r_p}{|\Delta v_\phi|_{x_s}} = \frac{4\pi}{3 \Omega_p x_s}$. Assuming the total libration period is dominated by traversing two such legs (one full cycle, neglecting the U-turn time), we estimate
\begin{equation}\label{eq:tlib}
t_{\text{lib}} \approx 2 t_{\text{leg}} \approx \frac{8 \pi}{3 \Omega_p x_s},
\end{equation}
such that $t_{\text{lib}} \approx \frac{4}{3 x_s} t_{orb}$, which yields $t_{\text{lib}}\approx 83$ orbits for our simulation parameters.

\subsection{Critical cooling time}

We can estimate the critical cooling parameter $\beta_{\text{crit}}$ required to prevent saturation of the entropy-related corotation torque by comparing the cooling timescale, $t_{\text{cool}}$, to the horseshoe libration timescale (Eq. \ref{eq:tlib}). Following the principle that the gradient-restoring mechanism (here, cooling) must operate on a timescale comparable to or shorter than the libration time to prevent saturation (\PK), the condition for de-saturation is $t_{\text{cool}} \lesssim t_{\text{lib}}$. Using our definition $t_{\text{cool}} = \beta / \Omega_p$, this condition becomes
\begin{equation}\label{eq:beta_crit_nl}
\beta_{\text{crit, \text{NL}}} \approx \frac{8 \pi}{3 x_s}.
\end{equation}
For the disk parameters used here ($x_s\approx 0.016$), we get $\beta_{\text{crit}} \approx 524$. Therefore, simulations with $\beta \lesssim 524$ are expected to experience largely unsaturated corotation torques due to thermal effects, assuming viscosity is also sufficient. 

In addition to the critical cooling time for desaturating the entropy-related horseshoe drag, we may define a second, shorter critical cooling parameter corresponding to the transition into the linear regime. This is motivated by the fact that, according to \PK, the nonlinear horseshoe drag persists as long as the diffusion (or cooling) timescale is shorter than the libration time, $t_{\rm cool} \lesssim t_{\rm lib}$. However, the torque approaches the linear regime if diffusion is fast enough to affect the horseshoe turn itself, which occurs on the much shorter U-turn time, $t_{\rm cool} \lesssim t_{\rm U-turn}$.

The U-turn time ($t_{\rm U-turn}$) represents the characteristic timescale on which the non-linear horseshoe drag develops and replaces the initial linear corotation torque after a planet is introduced into a disc. This timescale is not derived from first principles but is an empirical finding from hydrodynamic simulations. In their analysis, \citet{paardekooper2009a} demonstrated that this setup time is a fixed fraction of the libration period, observing from their simulation results that $t_{\rm U-turn} \approx 0.15 \ t_{\rm lib}$, which corresponds to $\frac{3}{20} t_{\rm lib}$.

Thus, the critical cooling parameter for the linear regime is
\begin{equation}\label{eq:beta_crit_lin}
\beta_{\rm crit,lin} \approx  \frac{2 \pi}{5 x_s }.
\end{equation}
For our fiducial value $x_s \approx 0.016$, this yields $\beta_{\rm crit,lin} \approx 80$. When $\beta \lesssim \beta_{\rm crit,lin}$, the corotation torque transitions to the linear regime, with a substantial reduction in magnitude compared to the nonlinear horseshoe drag. 

\subsection{Critical thermal diffusivity}
\PK, who considered thermal diffusion, defined the parameter 
\begin{equation}
p_\chi = \sqrt{\frac{r_p^2 \Omega_p x_s^3}{2 \pi \chi_p}},     
\end{equation}
which compares the thermal diffusion time across $x_s$, $t_{\chi} = (x_s r_p)^2/\chi_{p}$, to the libration time: $p_\chi^2 \propto t_{\chi} / t_{\text{lib}}$.

The condition $p_\chi=1$ defines a critical thermal diffusivity, $\chi_{\text{crit, NL}}$, below which the horseshoe drag saturates:
\begin{equation}\label{eq:chi_crit_nl}
\chi_{\text{crit, NL}} = \frac{1}{2\pi} x_s^3 r_p^2 \Omega_p.
\end{equation}
For our parameters ($x_s \approx 0.016$), this yields $\chi_{\text{crit, NL}} \approx 6.5 \times 10^{-7} r_p^2 \Omega_p$.

Similar to the cooling time, we can define a second critical diffusivity corresponding to the transition to the linear regime, occurring when the diffusion timescale becomes comparable to the U-turn timescale ($t_{\chi} \lesssim t_{\text{turn}} \approx \frac{3}{20}t_{\text{lib}}$). This requires a diffusivity that is a factor of $20/3$ larger than the saturation threshold:
\begin{equation}\label{eq:chi_crit_lin}
\chi_{\text{crit, lin}} \approx \frac{20}{3} \chi_{\text{crit, NL}} = \frac{10}{3\pi} x_s^3 r_p^2 \Omega_p.
\end{equation}
For our parameters, this yields $\chi_{\text{crit, lin}} \approx 4.3 \times 10^{-6} r_p^2 \Omega_p$ ($p_{\chi} \approx 0.4$). This suggests that for $\chi_p \gtrsim 4.3 \times 10^{-6}$, the entropy-related torque is expected to deviate from the fully non-linear prediction and approach its linear value.

\subsection{Critical viscosity}
In analogy to the above critical quantities, we can define critical values of viscosity. In the \PK~ model the role of viscosity is fundamentally characterized by the dimensionless parameter $p_\nu$:
\begin{equation}\label{eq:pnu}
  p_\nu = \frac{2}{3}\sqrt{\frac{r_{p}^{2}\Omega_{p}}{2\pi\nu_p} x_{s}^{3}}.
\end{equation}
This parameter is directly analogous to $p_\chi$ defined above. Thus,
    large $p_\nu\gtrsim 1$ implies that viscous diffusion is slow relative to libration, allowing phase mixing to erase gradients, leading to saturation.
    On the other hand, for small $p_\nu \ll 1$  viscous diffusion is fast relative to libration, which prevents saturation but also disrupts the non-linear mechanism, pushing the torque towards its linear limit (see Section \ref{sec:nu_lin}).
    
    However, in contrast to thermal diffusion, viscous diffusion is required to de-saturate \emph{both} the entropy-related as well as the vortensity-related horseshoe drags. Likewise, if viscosity is too strong, the entire corotation torque is pushed into the linear regime, whereas in absence of any viscosity, the \PK~ model predicts an identically vanishing final corotation torque, leaving only the Lindblad torque\footnote{Of course, the Lindblad torque is affected by viscosity as well, via the damping of density waves, but investigating this is beyond the scope of this paper.}.

The condition $p_\nu = 1$ defines a critical viscosity at which the horseshoe drag transitions from being mostly unsaturated to becoming saturated:
\begin{equation}
  \nu_{\text{crit, NL}} = \frac{2}{9\pi} x_s^3 r_p^2 \Omega_p.
  \label{eq:nucrit_final}
\end{equation}
For our parameters ($x_s \approx 0.016$), Eq.~\eqref{eq:nucrit_final} gives $\nu_{\text{crit, NL}} \approx 2.9 \times 10^{-7} r_p^2 \Omega_p$, or, equivalently $\alpha_{\text{crit, NL}} = \nu_{\text{crit, NL}}/(h_p^2 r_p^2 \Omega_p) \approx 1.2 \times 10^{-4}$. This relatively low viscosity is sufficient to keep the torque largely unsaturated.

\subsection{Transition to Linear Torque at High Viscosity}\label{sec:nu_lin}

While $p_\nu \lesssim 1$ prevents nonlinear saturation, if $p_\nu$ becomes very small (high viscosity), the torque transitions from its non-linear value towards its linear limit. 
This transition occurs when viscosity is strong enough to affect the horseshoe U-turn itself. \PK~ argue this happens when the viscous timescale $t_\nu$ across the horse shoe region becomes comparable to or shorter than the U-turn time, $t_{\text{turn}} \approx \frac{3}{20} t_{\text{lib}}$. For our parameters, $t_{\text{lib}} \approx 42 T_{\text{orb}}$, so $t_{\text{turn}} \approx 6.25 T_{\text{orb}}$. The condition $t_\nu \lesssim t_{\text{turn}}$ defines a higher critical viscosity $\nu_{\text{crit,turn}}$. Using the relation between $t_{\text{turn}}$ and $t_{\text{lib}}$ derived above, and scaling from the non-linear critical viscosity $\nu_{\text{crit,NL}}$ (Eq. \ref{eq:nucrit_final}), we can define:
\begin{equation}\label{eq:nu_crit_lin}
    \nu_{\text{crit,lin}} \approx \frac{20}{3} \nu_{\text{crit,NL}} = \frac{40}{27\pi} x_s^3 r_p^2 \Omega_p.
\end{equation}
For our parameters, this yields $\nu_{\text{crit,lin}} \approx 1.9 \times 10^{-6} r_p^2 \Omega_p$ ($\alpha_{ \text{crit,turn}} \approx 7.7 \times 10^{-4}$). This corresponds to $p_\nu \approx 0.39$. For viscosities around or above this value ($p_\nu \lesssim 0.4$), the non-linear horseshoe flow breaks down, and the torque approaches its linear value.

\section{Transition and Saturation Functions}\label{app:transitionfunctions}
Here we elaborate slightly more on the saturation and transition functions of the PK11-model described in Section \ref{sec:total_torque_formula}.

     The saturation function $F(p)$ accounts for the potential saturation of the non-linear horseshoe torque due to insufficient diffusion over the libration period. It arises from modelling the steady-state balance between advection (maintaining property contrasts) and diffusion (erasing them) over the libration timescale $t_{lib}$. 
     Derived from solving a simplified transport equation \citep{Masset2002}, $F(p)$ represents the fraction of the unsaturated torque that survives diffusion. 
     $F(p) \approx 1/(1+(p/1.3)^2)$ provides a good approximation (\PK); 
     $F(p) \approx 1$ for $p \lesssim 1$ indicates sufficient diffusion to prevent saturation, while $F(p) \to 0$ for $p \gg 1$ indicates strong saturation due to weak diffusion.

    The transition function $G(p)$ models the reduction of the non-linear horseshoe torque ($\Gamma_{HS}$) itself when diffusion becomes strong enough ($p \ll 1$) to interfere with the dynamics of the horseshoe U-turn. 
    It is derived by integrating the torque contribution over all streamlines within the horseshoe region (radial coordinate $x$ from $0$ to $x_s$). The contribution of each streamline is weighted by a factor $\mathcal{F}$ which quantifies its effectiveness in maintaining its property contrast (vortensity or entropy) during the U-turn in the presence of diffusion (\PK Eqs.~25--27). This factor $\mathcal{F}$ decreases smoothly from 1 towards 0 when the diffusion time becomes short compared to the characteristic U-turn time ($\tau_{turn} \approx \frac{3}{20} t_{lib}$). As viscosity increases ($p$ decreases), the weighting factor $\mathcal{F}$ begins to drop below 1 first for streamlines with the largest radial extent ($x \approx x_s$), causing their contribution to the non-linear torque to diminish. Thus, $G(p)$ represents the overall fraction of the full horse shoe torque $\Gamma_{HS}$ that remains operative when strong diffusion affects the U-turn process itself.

 The transition Function $K(p)$ governs the emergence of the linear corotation torque ($\Gamma_{lin}$) as the non-linear horseshoe torque fades at high diffusion ($p \ll 1$). It ensures that the total torque smoothly transitions to the correct linear value as $p \to 0$ (where $G(p) \to 0$ and $K(p) \to 0$, leaving the $(1-K)$ term dominant). \PK derived $K(p)$ using the same functional form as $G(p)$ (as both arise from diffusion affecting streamline dynamics) but parameterized it with a different characteristic transition scale ($p_{crit}$ or $\tau_0$). This allows the fading of the non-linear torque and the emergence of the linear torque to occur at slightly different rates as $p$ varies, providing a better fit to simulation results across the transition, motivated by the potentially different timescales governing horseshoe dynamics versus linear wave interactions near resonance (\PK).

\section{Vortex Identification and Mass Calculation}
\label{app:vortex_mass}

To obtain an estimate for the mass of the relevant COS-induced vortices that form in our planet-disk interaction simulations, we employ a multi-step post-processing analysis. The procedure is as follows:

\begin{enumerate}
    \item \textbf{Vorticity Perturbation Map:} We first compute the vertical component of the vorticity, $\omega_z$, from the gas velocity fields $(v_r, v_\phi)$ in our 2D cylindrical coordinate system $(r, \phi)$:
    \begin{equation}
        \omega_z = \frac{1}{r}\frac{\partial (r v_\phi)}{\partial r} - \frac{1}{r}\frac{\partial v_r}{\partial \phi}.
    \end{equation}
    To isolate the non-axisymmetric vortex structures from the background Keplerian shear, we calculate the vorticity perturbation, $\delta\omega_z = \omega_z(t) - \omega_z(t=0)$, by subtracting the initial axisymmetric vorticity field from that of the simulation snapshot. The resulting 3D field is then vertically averaged to produce a 2D map, $\langle\delta\omega_z\rangle_z(r, \phi)$.

    \item \textbf{Vortex Identification:} Anticyclonic vortices appear as regions of local minima in the $\langle\delta\omega_z\rangle_z$ map. To robustly identify the centers of these vortices while excluding spurious detections from small-scale turbulence, we first smooth the vorticity map with a 2D Gaussian filter. The standard deviation of the filter is chosen to correspond to a physical scale of $0.5 H_0$. This ensures that only coherent structures with a radial extent of at least this scale are considered. We then use a peak-finding algorithm (\texttt{skimage.feature.peak\_local\_max}) on the negative of this smoothed map to locate the vortex centers.

    \item \textbf{Mass Calculation:} For each identified vortex center, we define its boundary using the original, unsmoothed vorticity map. The boundary is defined as the contour level corresponding to 50\% of the peak vorticity value at the vortex center. The largest closed contour at this level that encloses the center is taken to be the vortex area, $A_v$. We compute the surface density perturbation, $\delta\Sigma(r, \phi) = \Sigma(r, \phi) - \Sigma_0(r)$, where $\Sigma_0(r)$ is the azimuthally averaged initial surface density profile. The mass of the vortex, $M_v$, is then calculated by integrating this surface density perturbation over the vortex area:
    \begin{equation}
        M_v = \iint_{A_v} \delta\Sigma(r, \phi) \, r \, dr \, d\phi.
    \end{equation}
\end{enumerate}

\begin{figure*}
 \centering 
 	\includegraphics[width=\textwidth]{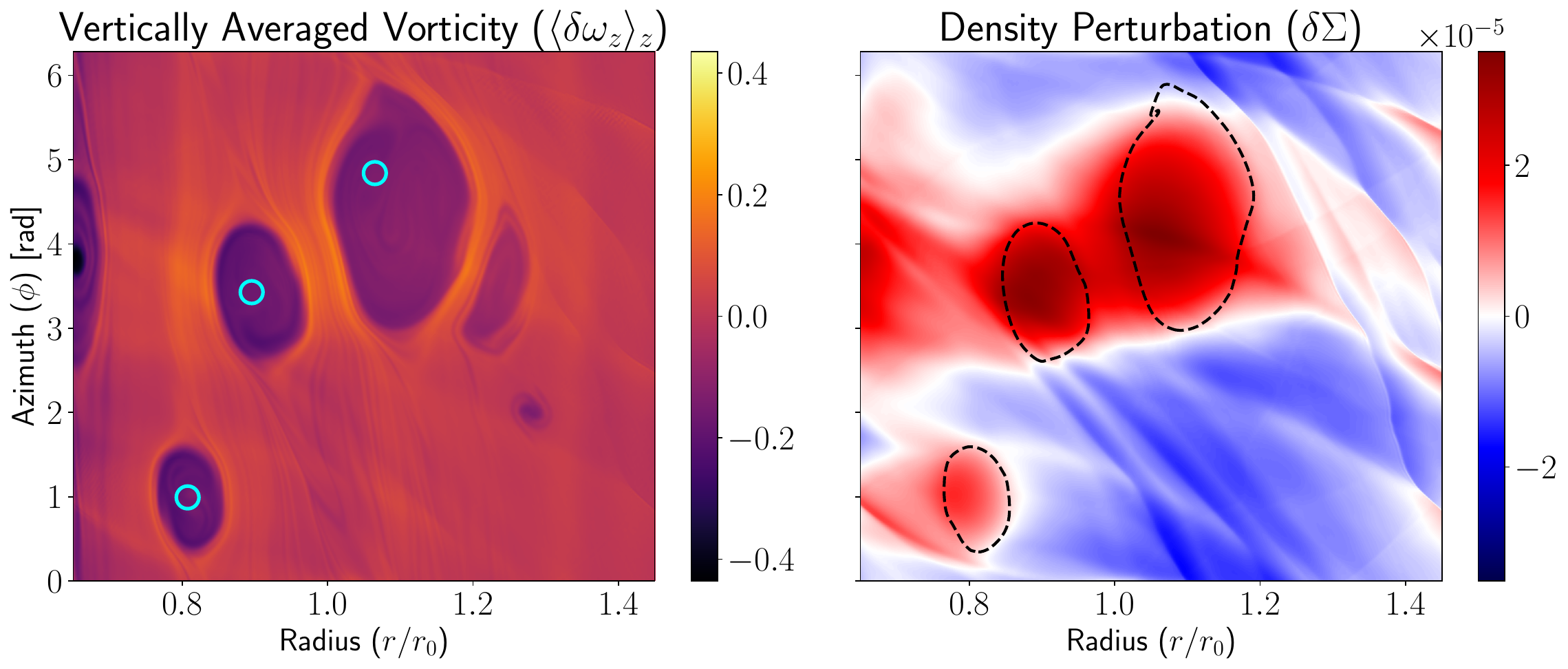}
     \caption{Vertically averaged $z$-component of vorticity (left) and density perturbation against the initial density profile. In this image the algorithm described in the text has identified three vortices. The vortex at radius $r\sim 1.1$ is about to interact with the planet at $r=1$ to provide a positive torque "kick".}
     \label{fig:vortex_mass}
 \end{figure*}

Using this method, we identify three primary vortices in the 3D simulation with thermal diffusion method $A$ and $\chi_p = 2\cdot 10^{-5}$ at 664 orbits, as shown in figure \ref{fig:vortex_mass}. Their calculated masses are on the order of $10^{-6}$ to $10^{-5}$ in code units. This confirms that the vortices, while significant drivers of local dynamics, are substantially less massive than the embedded planet ($M_p = 2.25 \times 10^{-5}$). In particular, the large vortex that passes by the planet around 700 orbits and provides a positive torque kick has a mass of $8.1\cdot 10^{-6}$, which is $0.36 M_p$. Most of the vortices are less massive than this one.

\end{document}